\RequirePackage{ifpdf}
\documentclass[paper]{JHEP3}
\pdfoutput=1
\usepackage{epsfig}
\usepackage{comment}
\usepackage{amsmath}

\newcommand{\roughly}[1]{\mathrel{\raise.3ex\hbox{$#1$\kern-0.85em
\lower1ex\hbox{$\sim$}}}}

\def\endignore{}
\def\ignore #1\endignore{} 

\def\subsubsubsection#1{{\bigskip\noindent{\em #1:}}}

\def\cA{{\cal A}}
\def\cB{{\cal B}}
\def\cC{{\cal C}}
\def\cD{{\cal D}}
\def\cH{{\cal H}}

\def\cM{{\cal M}}
\def\cN{{\cal N}}
\def\cO{{\cal O}}
\def\cP{{\cal P}}
\def\cR{{\cal R}}

\def\cW{{\cal W}}

\def\ss{\scriptscriptstyle}
\def\ssA{{\ss A}}
\def\ssB{{\ss B}}

\def\talpha{{\omega}}

\def\ol#1{\overline{#1}}
\def\Dsl{\hbox{/\kern-.6700em\it D}} 
\def\dsl{\hbox{/\kern-.5300em$\partial$}}

\def\eqa{\begin{eqnarray}}
\def\eeqa{\end{eqnarray}}
\def\eq{\begin{equation}}
\def\eeq{\end{equation}}
\def\be{\begin{equation}}
\def\ee{\end{equation}}
\def\bea{\begin{eqnarray}}
\def\eea{\end{eqnarray}}
\def\nn{\nonumber}

\def\pref#1{(\ref{#1})}

\def\Tr{{\rm Tr \,}}
\def\exd{{\rm d}}

\title{EFT Beyond the Horizon: 
Stochastic Inflation and How Primordial Quantum Fluctuations Go Classical}
\author{C.P.~Burgess,$^{1,2,3}$ R.~Holman,${}^{4}$ G.~Tasinato${}^5$ and M. Williams$^{1,2}$ \\
$^1$ Physics \& Astronomy, McMaster University,
Hamilton, ON, Canada, L8S 4M1\\
$^2$ Perimeter Institute for Theoretical Physics, Waterloo, ON, Canada N2L 2Y5\\
${}^3$ Division PH\,-TH, CERN, CH-1211, Gen\`eve 23, Suisse.\\
$^4$ Physics Department, Carnegie Mellon University, Pittsburgh PA
15213 USA\\
$^5$ ICG, University of Portsmouth,
 Portsmouth, United Kingdom, PO1 3FX.
}

\abstract{We identify the effective field theory describing the physics of super-Hubble scales and show it to be a special case of a class of effective field theories appropriate to open systems. Open systems are those that allow information to be exchanged between the degrees of freedom of interest and those that are integrated out, such as would be appropriate for particles moving through a fluid. Strictly speaking they cannot in general be described by an effective lagrangian; rather the appropriate `low-energy' limit is instead a Lindblad equation describing the time-evolution of the density matrix of the slow degrees of freedom. We derive the equation relevant to super-Hubble modes of quantum fields in de Sitter (and near-de Sitter) spacetimes and derive two of its implications. We show that the evolution of the diagonal density-matrix elements quickly approach the Fokker-Planck equation of Starobinsky's stochastic inflationary picture. This allows us both to identify the leading corrections and provide an alternative first-principles derivation of this picture's stochastic noise and drift. (As an application we show how the noise changes for systems with a sub-luminal speed of sound, $c_s < 1$.) We then argue that the presence of interactions drive the off-diagonal density-matrix elements to zero in the field basis. This shows why the field basis is generally the `pointer basis' for the process that decoheres primordial quantum fluctuations while they are outside the horizon, thus allowing them to re-enter later as classical field fluctuations, as assumed when analyzing CMB data. The decoherence process is very efficient, occurring after several Hubble times even for interactions as weak as gravitational-strength. Crucially, the details of the interactions largely control only the decoherence time and not the nature of the final late-time stochastic state, much as interactions can control the equilibration time for thermal systems but are largely irrelevant to the properties of the resulting equilibrium state.
}

\preprint{CERN-PH-TH-2014-142}

\begin{document}

\makeatletter \@addtoreset{equation}{section} \makeatother
\renewcommand{\theequation}{\thesection.\arabic{equation}}

\setcounter{page}{1} \pagestyle{plain}
\renewcommand{\thefootnote}{\arabic{footnote}}
\setcounter{footnote}{0}

\section{Introduction}

The advent of precision CMB cosmology reveals the Universe to be a somewhat lumpy place whose present crags and wrinkles partly reflect an earlier accelerated lifestyle. Cosmologists infer properties of this earlier epoch much as one might try to guess about past excesses by gazing on the features of one long past the sowing of wild oats.

In particular, evidence continues to build that the right explanation for present-epoch super-Hubble correlations lies with quantum fluctuations generated during a much-earlier epoch of accelerated expansion. A common feature of such explanations is that quantum fluctuations pass to super-Hubble scales in the remote past and then re-enter as classical fluctuations after spending a lengthy period frozen beyond the Hubble pale. This kind of picture raises two related and oft-considered issues:
\begin{enumerate}
\item {\em What effective theory describes long-wavelength physics in the super-Hubble regime?}
\item {\em Why do quantum fluctuations re-enter the Hubble scale as classical distributions?}
\end{enumerate}

The first of these issues starts with the observation that for most physical systems the long-wavelength limit is usually most efficiently described by a Wilsonian effective field theory (EFT), obtained by integrating out shorter-wavelength modes \cite{EFTs}. Since super-Hubble modes have the longest wavelengths of all, one is led to ask what field theory provides its effective description.\footnote{Notice this is distinct from other EFTs used for inflation, such as those that express the decoupling of short-distance modes during inflation \cite{TPI} and those describing the fluctuations about inflationary backgrounds \cite{InfEFT}.} Such an effective description might allow a cleaner understanding of the various thorny infrared issues faced by quantum fields on de Sitter space \cite{CMBth1}---\cite{CMBthn}.

The second issue asks why fluctuations that are initially described (at horizon exit) in terms of vacuum correlators of field operators, $\langle \Phi(x) \Phi(y) \rangle$, eventually instead become interpretable (at horizon re-entry) in terms of an ensemble average of classical field configurations, $\varphi(x)$.

Because they have been oft-considered, there is also a party line about both of these issues in the cosmology community. It states that Starobinsky's stochastic formulation of inflation \cite{starobinsky} provides the answer to the first question. The formulation is stochastic because it reflects the fact that perturbations begin to evolve independently once their wavelengths are stretched outside the Hubble scale, because the strong red-shifting suppresses the energy cost of such long-wavelength gradients. This allows the field value to drift from one Hubble patch to another over time in what is essentially a random walk. Consequently, for such scales, it is more instructive to track the probability distribution, $P[ \phi \,]$, for the values taken by slowly-varying fields, $\phi(x)$, coarse-grained over each Hubble patch, than to follow its detailed evolution for a specific set of classical initial conditions.

The party line for the second issue is that it need not be solved at all (see, however, \cite{NonPLdecoherence}). The reason is because inflationary systems exhibit `decoherence without decoherence' \cite{starobinsky, Guth:1985ya, Polarski:1995jg}. This delightfully named phenomenon is based on the property that quantum states become `squeezed' during their sejour outside the Hubble scale, in such a way that each mode's canonical momentum rapidly shrinks to zero over several Hubble times \cite{squeezed}. This implies that only the expectation values of fields (rather than their canonical momenta) remain unsuppressed at late times,
\bea
 \langle \Phi(x_1) \dots \Phi(x_n) \rangle &=& \int \cD \phi \Bigl[ \phi(x_1)\dots \phi(x_n) \Bigr] \langle \phi(\cdot) |\Xi|\phi (\cdot) \rangle \nn\\
 &=& \int \cD \phi \Bigl[ \phi(x_1)\dots \phi(x_n) \Bigr] P[\phi\,] \,,
\eea
and because these only require the diagonal elements of the density matrix, $\Xi$, it is irrelevant what its off-diagonal elements do.

In this paper we argue that although the party line \cite{starobinsky, OtherNoiseDeriv} is essentially correct, as far as it goes, more can be said about these two questions by embedding them into the language of effective field theories. In particular, doing so potentially allows a more systematic derivation of the corrections to the leading stochastic formulation. We also argue that the two questions above are related to one another. In particular, although it is true that squeezing ensures that off-diagonal terms in the density matrix need not be tracked when computing correlation functions, we argue that a complete effective description nonetheless does permit them to be tracked. Furthermore, such tracking reveals that they  rapidly fall to zero. Conceptually, it is not merely that primordial fluctuations behave {\em as if}  they were classical; in fact they actually rapidly do become classical with a decoherence time-scale similar to the squeezing time-scale. In this sense the answer to the first question also provides a more direct answer to the second.

The effective theory we present here for super-Hubble physics turns out to be a special case of how EFTs can be formulated for a more general class of physical systems for which the long-wavelength physics {\em cannot} be captured by some sort of effective lagrangian \cite{Companion}. An effective lagrangian does not exist because --- unlike for most long-wavelength systems usually encountered in EFT applications --- the degrees of freedom integrated out to obtain the long-distance theory are not excluded by a conservation law. As a result the long-distance degrees of freedom can exchange information with the short-distance ones, making the effective theory for super-Hubble physics more like the effective theory for a particle moving through a fluid than a traditional low-energy EFT. In particular, because information can be exchanged in open systems like these, pure states can evolve into mixed states, and this implies the time-evolution cannot be described by hamiltonian evolution.

This should be contrasted with a traditional low-energy EFT \cite{EFTs}, for which energy conservation precludes high-energy states from turning up at late times if not initially present. Because the low-energy fields form a basis of operators, in the usual formulation they can always be used to describe {\em any} time evolution within the low-energy regime --- including the influence of virtual heavy states. This ultimately underlies the existence of a low-energy effective hamiltonian (or lagrangian) cast purely in terms of the light degrees of freedom.

Of course none of this means that there is no simplification to be had by exploiting hierarchies of scales in open systems. In particular, given that energy is not used to identify the states for integrating out, it is more useful to track the distinction between `slow' and `fast' processes. In particular, great simplifications occur if the physics of interest takes place on time scales that are long compared with the correlation times of the environment through which it moves. In this case the environment can forget its entanglement with the degrees of freedom being followed, allowing the late-time evolution to be described in a controlled way as a Markov process. Such a formulation can often allow a systematic identification of late-time evolution that resums the secular growth that is sometimes present over the shorter time-scales for which perturbation theory is directly valid.

The appropriate language to express this control is the Lindblad equation \cite{Lindblad}, which describes the coarse-grained time-evolution of the density matrix of a small subsystem over times long compared with the environmental correlation time, $\tau$, but short compared to the times over which perturbative methods break down (and so for which secular growth can become important) \cite{Companion}. That is, if the interaction energy between the subsystem and environment is of order $V$, then the Lindblad equation allows the coarse-grained time evolution to be developed in systematic powers of $V\tau$.

We here derive the Lindblad equation appropriate for the evolution of super-Hubble modes of a scalar field (on curved spacetime) once the physics of sub-Hubble modes is integrated out. For applications to cosmologies with accelerated expansion we coarse-grain over short-wavelength sub-Hubble modes, which thereby serve as the environment through which long-wavelength super-Hubble modes evolve. Our treatment follows closely that of \cite{StochInfOld}, and differs in detail from (but is very similar in spirit to) the approach taken in \cite{DGLAP}. In this case the environmental correlation time is the Hubble time, $H^{-1}$, itself, and so the simplified controlled treatment of slow degrees of freedom necessarily only applies to modes well outside the Hubble scale, $k/a \ll H$. See also \cite{Kiefer:2006je} for an application of the Lindblad equation to the problem we are considering.

If the interaction energy between super-Hubble and sub-Hubble modes is of order $V$, then the Lindblad equation allows the coarse-grained time evolution of super-Hubble modes to be developed systematically in powers of $V/H$. To linear order in $V/H$ the only effect of these interactions is to provide a mean-field modification of the effective hamiltonian for the evolution of super-Hubble modes.

It is at second order in $V/H$ that the effects of interactions can be seen to cause pure-to-mixed evolution and to decohere the super-Hubble modes (in the sense that the density matrix, $\Xi$, for super-Hubble modes rapidly becomes diagonal). The `pointer' basis, in which $\Xi$ becomes diagonal, is generically the one that diagonalizes the interactions $V$, but for super-Hubble modes this is always the field basis because the squeezing of modes ensures all canonical momenta in $V$ are rapidly driven to zero.\footnote{The arguments we make here closely follow those of ref.~\cite{StochInfOld}, which applied the same formalism to study decoherence on much shorter scales during reheating. Our conclusions differ from this reference inasmuch as some of us argued at that time that it would be unlikely that the techniques used here could decide whether decoherence could occur during inflation --- as had been suggested to us at the time as being the probable picture by Robert Brandenberger. At that time we did not understand how to arrange a hierarchy of time-scales in a controlled way over super-Hubble scales.} The upshot is that after relatively few Hubble times the full density matrix, $\langle \psi(\cdot) | \Xi | \phi(\cdot) \rangle$, can be replaced with the classical probabilities, $P[\phi] = \langle \phi(\cdot)| \Xi | \phi(\cdot) \rangle$.

Finally, in the absence of interactions (that is, when $V = 0$, so we just have free fields propagating in a cosmological spacetime) the Lindblad equation reduces to the usual Liouville equation for the free evolution of the reduced density matrix. We use this equation to derive the evolution equation for the diagonal elements, $P[\phi\,]$, and show --- {\em separately, for each mode} --- that these satisfy a Fokker-Planck equation, including a noise and drift term.\footnote{To capture the drift term we track the adiabatic influence of interactions at long wavelength at leading order in slow-roll parameters.} We compute how the noise term evolves with time and show that it vanishes while the mode in question is inside the Hubble scale, but becomes nonzero once the mode exits.

Physically, the noise arises because each mode evolves as an independent harmonic oscillator, and at horizon exit the oscillator potential flips from being stable (concave up) to unstable (concave down). As a result an initially gaussian ground state converts to an oscillatory state that is not localized in field space. The noise term in the Fokker-Planck equation vanishes for the stable oscillator, but becomes nonzero once the barrier is removed and the oscillator is allowed to explore values away from the potential's stationary point.

For free fields decoherence does not occur (which in this limit agrees with the party line), and canonical momenta get squeezed to zero. But the kinetic term in the energy converts quantum fluctuations into stochastic noise on a mode-by-mode basis. The total noise for coarse-grained fields in position space receives contributions from both the mode-by-mode noise and the coarse graining itself, with the sum reproducing standard calculations.

Although the noise discussion is performed for free particles in de Sitter space, we also relax this slightly to include the leading new terms in a slow-roll expansion for the long-wavelength interactions associated with a potential $V(\phi)$. This allows us also to compute the drift contribution to the Fokker-Planck equation, which again agrees with standard results.

We present these ideas as follows. \S\ref{sec:OpenEFT} starts by briefly reviewing the Lindblad formalism for open systems, along the lines of \cite{Companion}. \S\ref{sec:FreeFields} then translates standard results for free fields in de Sitter (and near-de Sitter) geometries into a density-matrix language, using them to solve the Liouville equation for the time-evolution of the density matrix. In particular we show how the Liouville equation for the diagonal elements of the density matrix can be written as a Fokker-Planck equation and evaluate the time-dependence of the resulting noise kernel for each mode. This section also includes interactions in a slow-roll expansion to obtain the drift contribution to the Fokker-Planck equation. Then we continue  with \S\ref{sec:Interactions} where interactions  play a significant role, and we show, closely following the steps of ref.~\cite{StochInfOld}, how they act to decohere the density matrix, and that  a robust estimate for the decoherence time-scale is set by a few Hubble times, regardless of the details of the form of the interaction between long- and short-wavelength modes. We finally summarize our arguments in \S\ref{sec:concl}, and speculate briefly on how similar arguments may help understand information-loss puzzles in black holes.

\section{Open EFTs}
\label{sec:OpenEFT}

In this section we briefly describe the Lindblad formalism \cite{Lindblad}, and its relevance to the effective field theory of open systems \cite{Companion}. Our presentation of the formalism itself follows closely that of ref.~\cite{BM}, but see also \cite{Feynman,others} for alternative discussions of the same formalism, and in particular \cite{Kiefer:2006je} for an application of the Lindblad equation to the problem we are considering.

\subsection{A hierarchy of scales for open systems} \label{subsec:openhier}

Our interest is in when a system's Hilbert space can be written as the direct product of an observable sector, $A$, and its environment, $B$: ${\cal S} = {\cal S}_A \otimes {\cal S}_B$. What is important is that we choose only to follow observables in sector $A$ and largely ignore those involving sector $B$.

This framework includes garden-variety low-energy effective theories, if we imagine $A$ to be the Fock space built using modes of the light particles making up the low-energy theory, while $B$ consists of the Fock states representing the heavy particles that are integrated out. In this case the choice to ignore $B$ observables amounts to the choice only to follow observables involving low-energy states.

But, crucially, this framework also goes beyond the low-energy set-up to include much more general situations where sectors $A$ and $B$ are not distinguished by energy, and so states can nontrivially evolve between sectors $A$ and $B$. An example of this type might be where $A$ represents the states of a particle moving through a fluid described by $B$. In this more general case the option not to follow observables in the $B$ sector might just be a convenient choice and not something dictated by a low-energy imperative.

As ever, the goal is to identify how quantities evolve in time, and to do so we write the total Hamiltonian governing time-evolution of system and environment as
\be
 \cH = \cH_\ssA + \cH_\ssB + V \,,
\ee
where $\cH_\ssA$ and $\cH_\ssB$ describe the separate dynamics of $A$ and $B$ and $V$ describes their mutual interactions. The interaction $V$ defines a time-scale, $\tau_p$, through the condition $V \tau_p \sim \cO(1)$, beyond which it is not straightforward to evolve systems perturbatively in $V$.

\smallskip

The principle of no free lunch ensures that in general the time-evolution of such a system is complex, with the interactions ensuring that potentially complicated correlations can build up between $A$ and $B$, even if these are not present initially --- that is, even if the initial system density matrix satisfies $\rho(t = t_{0}) = \varrho_A \otimes \varrho_B$. As is often the case, however, a great simplification arises if there is a hierarchy of scales, and in this case the simplification arises if \cite{Lindblad}:
\begin{enumerate}
\item The correlation time, $\tau_c$, over which the autocorrelations $\langle \delta V(t) \, \delta V(t - \tau_c) \rangle_B$ are appreciably nonzero for the fluctuations of $V$ in sector $B$, is sufficiently short i.e. $\tau_c \ll \tau_p$ (or, equivalently, $V \tau_c \ll 1$);
\item System $B$ is large enough not to be appreciably perturbed by the presence of system $A$.
\end{enumerate}

Such a hierarchy hands one a powerful theoretical tool since it ensures two things. First, it ensures that to good approximation system $B$ just sits there. Second, it ensures that the dynamics of system $A$ does not retain any memory of its correlations with $B$ over times $t \gg \tau_c$, and this allows the evolution of $A$ in the presence of $B$ to be treated as a Markov process. Furthermore, assumption 1 above implies this Markov evolution can be tracked perturbatively in $V$, at least for times $\tau_c \ll t \ll \tau_p$.

But these assumptions also open the way to computing evolution over times $t \gg \tau_p$, and this is where the Lindblad equation comes into its own. The procedure is to coarse-grain the time-derivative for $\rho_\ssA$ over times, $\Delta t$, satisfying $\tau_c \ll \Delta t \ll \tau_p$,
\be \label{coarsegrain}
 \left( \frac{\partial \rho_\ssA}{\partial t} \right)_{{\rm coarse} \atop {\rm grained}} = \frac{\Delta \rho_\ssA}{\Delta t} = F[V, \rho_\ssA(t)] \,,
\ee
for some (possibly complicated) function $F$. It is the Markov nature of the evolution that allows the right-hand-side to be written using only $\rho_\ssA(t)$ (and not its convolution over all times in the past of $t$). Furthermore, assumption 2 allows one to ignore the evolution of the properties of sector $B$ on the right-hand side, leaving $F$ only depending on $\rho_\ssA$. Finally, because the coarse graining is done over times small compared with $\tau_p$, the function $F$ can be computed in powers of $V$.

What is crucial is that this coarse graining could be done for {\em any} time, $t$, and so eq.~\pref{coarsegrain} applies equally well for a window of time, $\Delta t$, around any $t$. Only $\Delta t$, and not also $t$, must be small compared with $\tau_p$. Consequently the solutions to \pref{coarsegrain} can be trusted even for $t \gg \tau_p$. This makes the Lindblad formalism a natural one for resumming any secular time-dependence that may arise within perturbative evolution.

\subsubsection*{Applications to super-Hubble cosmology}

One of the main points of this paper is that the above framework provides the natural set-up for an effective treatment of super-Hubble physics, particularly within an accelerated cosmology.

In this case system $A$ represents the super-Hubble modes of all light fields for which $M^2 := (k/a)^2 + m^2 \ll H^2$. System $B$ represents all of the shorter-wavelength and more massive modes. Although the long-wavelength limit is in spirit a low-energy limit, the time-dependence of the scale factor, $a(t)$, ensures it has an important differences from standard low-energy limits. In particular, even free modes transit over time from system $B$ to system $A$, and in the traditional derivation it is this continual trickle of modes from $B$ to $A$ that is the source of stochastic noise.

System $B$ also has a natural correlation time in accelerating cosmologies: the inverse of the Hubble scale $\tau_c \sim H^{-1}$. This provides a natural correlation time because causality ensures fields in different Hubble patches evolve from one another in an uncorrelated way. This exchange of information between $B$ and $A$ is what makes it difficult to apply standard EFT techniques leading to an effective lagrangian describing super-Hubble physics.

Secular evolution and difficulties in inferring late-time behaviour perturbatively also arise in cosmological contexts, to which the Lindblad formalism might naturally be expected to apply. In particular, its solutions may provide a good way to resum the secular evolution that is known to plague de Sitter (and near-de Sitter) cosmologies in particular (see also \cite{RS,deSRG,DGLAP} for other approaches to resumming secular evolution in de Sitter space).

\subsection{The Lindblad equation}

To compute $F$ (on the right-hand side of \pref{coarsegrain}) perturbatively in $V$, it is convenient to work within the interaction representation, for which the time evolution of operators is governed by $\cH_\ssA + \cH_\ssB$ while the evolution of states is governed by $V$. We concentrate on obtaining an evolution equation for the reduced density matrix, $\rho_A(t) = \Tr_B[\rho(t)]$, since this includes all of the information concerning measurements involving observables only in sector $A$.

To proceed we write the interactions as a sum of products of operators acting in each sector,
\be \label{Vdef}
    V(t) = \cA_i(t) \, \cB^i(t) \,,
\ee
where there is an implied sum\footnote{For local theories this implied sum could also include an integration over spatial position.} on $i$, ${\cal A}_i$ denotes a functional of the fields describing the $A$ degrees of freedom, and ${\cal B}^i$ plays a similar role for sector $B$.

In the special case that the degrees of freedom in sector $B$ have
very short correlation in time, $\tau_c$, their influence on the coarse-grained evolution, $(\partial \rho_A/\partial t)_{\rm c\,g}$, can be represented in terms of Markovian  interactions. That is, suppose
\be \label{shortdistcorrs}
    \langle \delta {\cal B}^i(t) \, \delta {\cal B}^j(t')
    \rangle_\ssB = \cW^{ij}(t) \,
    \delta(t-t') \,,
\ee
where $\langle \cdots \rangle_\ssB$ denotes an average over only the $B$ sector of the Hilbert space, $\delta {\cal B}^i(t) := {\cal B}^i(t) - \langle {\cal B}^i(t) \rangle_\ssB$, and $(\cW^{ij})^* = {\cal W}^{ji}$ is a calculable function that is of order $\tau_c$. Under the above assumptions the coarse-grained evolution equation becomes
\be \label{dmeqbody}
    \left( \frac{\partial \rho_\ssA}{\partial t} \right)_{\rm c\,g} =
    i  
  \; \Bigl[ \rho_\ssA , \cA_j
    \Bigr] \, \langle \cB^j
    \rangle_\ssB
    -  \frac12 
    \; \cW^{jk} \Bigl[
    \cA_j \cA_k \rho_\ssA + \rho_\ssA \cA_j \cA_k
    - 2 \cA_k \rho_\ssA \cA_j \Bigr]  \,,
\ee
up to second order in $V\tau_c = \tau_c/\tau_p$. Notice that this equation trivially implies $\partial\rho_\ssA /\partial t = 0$ for any $\rho_\ssA$ that commutes with all of the $\cA_j$. It is this equation whose solutions we seek for cosmological situations below.

\section{Scalar fields on de Sitter space}
\label{sec:FreeFields}

As a practical application of open EFTs, we start with a scalar field, $\chi(x)$, which we imagine {\em not} to be the inflaton, but rather to be a spectator whose energy density is not responsible for inflation. (We may drop this assumption later, since much of what we say also applies to the Sasaki-Mukhanov field  \cite{Mukhanov}.) We also start by neglecting all slow-roll suppressed quantities, effectively working in de Sitter space. The leading slow-roll corrections are analyzed in subsection \ref{sec-drift}.

We take the lagrangian for $\chi(x)$ to be
\bea \label{Ldef}
 L &=& - \frac12 \int \exd^3 x \sqrt{-g} \; \Bigl[ (\partial \chi)^2 + m^2 \, \chi^2 \Bigr] \nn\\
 &=& \sum_k  a^3 \Bigl( \dot\chi_k^* \dot \chi_k - M_k^2 \;\chi_k^* \chi_k \Bigr) \,,
\eea
where we have used the FRW metric in cosmic time,
\be
 \exd s^2 = - \exd t^2 + a^2(t) \delta_{ij} \, \exd x^i \exd x^j \,,
\ee
with $a = a_0 e^{H(t-t_0)}$ and $\dot H = 0$. To arrive at the second equality, we expand the field in box-normalized Fourier modes, $\chi_k^* = \chi_{-k}$. Finally, $M_k^2$ denotes the quantity
\be \label{defomk}
 M_k^2 = \frac{k^2}{a^2} + m^2 \,,
\ee
where $m$ is the particle mass.

We next express the system in canonical variables and solve for the vacuum wave-functional. Since each mode is independent we suppress (temporarily) the mode label $k$. The canonical momenta implied by the mode lagrangian, $L_k$, of \pref{Ldef} are
\be
 \Pi := \frac{\partial L}{\partial \dot \chi} = a^3 \dot \chi^*
 \quad \hbox{and} \quad
 \Pi^* := \frac{\partial L}{\partial \dot \chi^*} = a^3 \dot \chi \,,
\ee
and so the hamiltonian becomes
\be
 \cH = \Pi \dot \chi + \Pi^* \dot \chi^* - L
 = \frac{\Pi^* \Pi}{a^3} + a^3  M^2 \chi^* \chi \,.
\ee

\subsection{Time dependence}

Since we are neglecting interactions at this point, different modes propagate independently of one another and the wave-functional for the system factorizes into a product of a wave function for each mode. Working in the Schr\"odinger representation each mode satisfies the Schr\"odinger equation,
\be
 i \frac{\partial}{\partial t} \, \Psi[\chi] = \cH \,  \Psi[\chi]
 = \left[ - \frac{1}{a^3} \frac{\partial^2}{\partial \chi \, \partial \chi^*} + a^3 \, M^2 \chi^* \chi \right] \Psi[\chi] \,.
\ee
To solve this, we use a Gaussian ansatz:
\be \label{Gansatz}
   \Psi[\chi] := N \, \exp \Bigl( - a^3 \; \talpha \, \chi^* \chi \Bigr) \,,
\ee
from which we omit any linear terms because of the symmetry $\chi \to - \chi$ (which also ensures that $\langle \chi \rangle = 0$). Substituting into the Schr\"odinger equation leads to
\be
 i\left[ \frac{\dot N}{N} - a^3 \, \chi^* \chi \Bigl( \dot \talpha + 3H \talpha \Bigr) \right]   \Psi = \Bigl[ - \Bigl( - \talpha + a^3 \, \talpha^2 \chi^* \chi \Bigr) + a^3 \, M^2 \chi^* \chi \Bigr]   \Psi \,,
\ee
so equating coefficients of each of the powers of $\chi$ on both sides gives evolution equations for $\talpha$ and $N$:
\be \label{talphaeq}
 \dot \talpha + 3 H \talpha = -i  \talpha^2 +i M^2 \,,
\ee
and
\be \label{tNeq}
 \dot N = -i\talpha \, N\,.
\ee

The first of these equations can be solved explicitly for de Sitter geometries, as we now sketch (and is done more explicitly in Appendix \ref{app:solving}). In the remote past, when $M^2 \simeq (k/a)^2 \gg H M$, inspection of eq.~\pref{talphaeq} shows that the time-independent (vacuum) solution is given by $\talpha \simeq M \simeq k/a$. We use this below as the boundary condition in the remote past.

Changing variables from $t$ to $a = e^{Ht}$, and denoting by primes derivatives with respect to $a$, one finds $\talpha(a) = -i a H \, w'(a)/w(a)$ where $w(a)$ is a de Sitter mode function, given explicitly in terms of Bessel functions,
\be \label{massivebessel}
 w(a) = x^{3/2} \Bigl[ c_1 J_\nu \left( x \right) + c_2 N_\nu \left( x \right) \Bigr] \,,
\ee
evaluated with argument $x = k/aH$ with order $\nu = \sqrt{\frac94 - (m/H)^2}$. The solution that ensures $\talpha$ has a positive real part (as required for a normalizable vacuum wave-functional), that approaches the flat-space value, $k/a$, in the remote past then is given by the Hankel function,
\be \label{massivew}
 w(a) \propto \left( \frac{k}{aH} \right)^{3/2} H_\nu^{(2)}\left( \frac{k}{aH} \right) \,.
\ee

In terms of the mode function, $w(a)$, we have
\be
 \talpha + \talpha^* = -iaH \left( \frac{ w^* w' - w w^{*'}}{|w|^2} \right)
 = - \frac{i}{a^3} \left[ \frac{\cW(w, w)}{|w|^2} \right] = \frac{1}{a^3 |w|^2}\,,
\ee
and
\be
 \talpha - \talpha^* = -iaH \left( \frac{ w^* w' + w w^{*'}}{|w|^2} \right)
 = -iaH \left[ \frac{\left( |w|^2 \right)'}{|w|^2} \right] \,,
\ee
where $\cW(f,g) := a^3 \left( f^* \dot g - g \dot{f}^* \right)$ defines the Wronskian of two mode functions, $f$ and $g$, and the last line uses the conventional mode normalization $\cW(w,w) = i$. Furthermore, using $\talpha = -i \dot w/w$ in eq.~\pref{tNeq} gives
\be
 \frac{\dot N}{N} = -i \talpha = - \frac{\dot w}{w} \,,
\ee
which shows that the product $N w$ is time-independent. In particular we see that
\be
 |N|^2 \propto \frac{1}{|w|^2} = a^3 \Bigl( \talpha + \talpha^* \Bigr) \,,
\ee
as is required for $N$ to normalize the wave-functional, eq.~\pref{Gansatz}, for all times. The variance has a simple expression in terms of $w$:
\bea
 \langle \chi^* \chi \rangle &=& |N|^2 \int_{-\infty}^{\infty}
 \exd \chi^* \exd \chi \; \chi^* \chi \; \exp\Bigl[ - a^3 (\talpha + \talpha^*) \chi^* \chi \Bigr] \nn\\
 &=& \frac{1}{a^3 (\talpha + \talpha^*)} = |w|^2 \,.
\eea

These expressions become very simple when $m/H \to 0$, for which case the relevant Bessel function reduces to an elementary function, giving the (normalized) massless mode function,
\be \label{masslessw}
 \left. w(a) \right._{m \to 0} = \frac{H}{\sqrt{2k^3}} \left( \frac{k}{aH} \right) \left( 1 - \frac{iaH}{k} \right) e^{-ik/aH} \,,
\ee
and so
\be \label{masslessomega}
 \left. \omega(a) \right._{m \to 0} =  \frac{k^3}{a \left[ k^2 + (aH)^2 \right]} + \frac{iH k^2}{k^2 + (aH)^2} \,,
\ee
while
\be \label{masslessvariance}
 \langle \chi^* \chi \rangle =  \frac{H^2}{2k^3} \left[ 1 + \left( \frac{k}{aH} \right)^2 \right] \,.
\ee
These reproduce many well-known results \cite{dSWF, dSProps}.

\subsection{Make some noise!}

We next turn to the evolution equation for the diagonal density matrix elements, $\langle \chi | \rho | \chi \rangle$, as dictated by the Schr\"odinger equation. We use here the state
\be
 \langle \xi |   \rho | \chi \rangle =   \Psi[\xi] \,   \Psi^*[\chi] \,,
\ee
whose time-dependence is determined above.

We start by writing out the Liouville equation, which expresses the implications of the Schr\"odinger equation for the time-evolution of the density matrix:
\bea
 \frac{\partial}{\partial t} \langle \xi |   \rho | \chi \rangle &=& \frac{\partial  \Psi[\xi]}{\partial t} \,   \Psi^*[\chi] +   \Psi[\xi] \, \frac{\partial \Psi^*[\chi]}{\partial t} \nn\\
 &=& \Bigl( -i \cH   \Psi [\xi] \Bigr)   \Psi^* [\chi] +   \Psi[\xi] \Bigl( -i \cH   \Psi [\chi] \Bigr)^* \nn\\
 &=& i \left[ \frac{1}{a^3} \left( \frac{\partial^2}{\partial \xi \partial \xi^*} - \frac{\partial^2}{\partial \chi \partial \chi^*} \right) + a^3 \cM^2 (\chi^* \chi - \xi^* \xi) \right] \langle \xi | \rho | \chi \rangle \,.
\eea
Our interest is to specialize this to its implications for the diagonal matrix elements, $  P[\chi] := \langle \chi |   \rho | \chi \rangle$, for which we find
\be
 \frac{\partial}{\partial t} \langle \chi |   \rho | \chi \rangle = \left[ \frac{i}{a^3} \left( \frac{\partial^2}{\partial \xi \partial \xi^*} - \frac{\partial^2}{\partial \chi \partial \chi^*} \right) \langle \xi |   \rho | \chi \rangle \right]_{\xi = \chi} \,.
\ee

To simplify this it is useful to state an identity, proven in Appendix \ref{app:gaussian}, true for any gaussian density matrix of the form considered to this point. This states that for any gaussian of the form
\be
 \langle \xi |   \rho | \chi \rangle = N^2 \, \exp \Bigl[ - a^3 \Bigl( \talpha \, \xi^* \xi + \talpha^* \chi^* \chi \Bigr) \Bigr] \,,
\ee
the following identity is true:
\be\label{simpform}
 \left[ \left( \frac{\partial^2}{\partial \xi \partial \xi^*} - \frac{\partial^2}{\partial \chi \partial \chi^*} \right) \langle \xi |   \rho | \chi \rangle \right]_{\xi = \chi}
  = \left( \frac{\talpha - \talpha^*}{\talpha + \talpha^*} \right) \frac{\partial^2 }{\partial \chi \partial \chi^*} \; \langle \chi | \rho | \chi \rangle \,,
\ee
from which we find the following evolution equation for $\langle \chi | \rho | \chi \rangle$:
\be \label{noisyFPv}
 \frac{\partial}{\partial t} \langle \chi |   \rho | \chi \rangle = \cN \; \frac{\partial^2 }{\partial \chi \partial \chi^*} \; \langle \chi |   \rho | \chi \rangle
\ee
where
\be \label{Nexp}
 \cN := \frac{i}{a^3} \left( \frac{\talpha - \talpha^*}{\talpha + \talpha^*} \right) \,.
\ee
Using in this expression the time-dependence for $\talpha(t)$ found above gives our final form for the evolution of the diagonal density-matrix elements.

Eq.~\pref{noisyFPv} has the form of a Fokker-Planck equation, with the right-hand side describing the effects of coupling to noise. (The generalization to include a `drift' term in this equation is straightforward  --- see \S\ref{sec-drift} for details.) Notice that this noise term relies for its presence on there being a nonzero imaginary part of $\talpha$, something that only becomes true on horizon exit, since $\talpha$ is real for the ground state of a garden-variety static harmonic oscillator in flat space. In the present instance this noise arises due to the residual influence of the quantum fluctuations (as described by the off-diagonal density-matrix elements) once they leave the Hubble horizon.

The time-dependence of $\cN$ is explicitly computed in Appendix \ref{app:solving}, and for a massless scalar field is given by
\be \label{masslessnoise}
 a^3\cN = - \left( \frac{ aH}{k} \right) \,.
\ee
The negative sign of this result is a general consequence of the freezing of the de Sitter mode functions for $k/a \ll H$, as can be seen from the general relation between $\cN$ and the time-dependence of the mode variance, $\langle \chi^* \chi \rangle$, implied by eq.~\pref{noisyFPv}:
\be
 \partial_t \langle \chi^* \chi \rangle = \int_{-\infty}^\infty \exd \chi \exd \chi^* \; \chi^* \chi \, \partial_t \langle \chi| \rho | \chi \rangle = \cN \int_{-\infty}^\infty \exd \chi \exd \chi^* \; \chi^* \chi \, \frac{\partial^2}{\partial \chi \partial \chi^*} \langle \chi| \rho | \chi \rangle = \cN \,,
\ee
where the last equality integrates by parts and uses the normalization condition Tr$\,\rho = 1$.

\subsubsection*{Combining momentum modes}

The previous section shows that each momentum mode satisfies a Fokker-Planck equation with noise,
\be \label{noisyFPk}
 \frac{\partial}{\partial t} \langle \chi_k |   \rho | \chi_k \rangle = \cN_k(t) \frac{\partial^2 }{\partial \chi_k \partial \chi_k^*} \; \langle \chi_k |   \rho | \chi_k \rangle \,,
\ee
where $\cN_k$ is given by eq.~\pref{Nexp}, reducing to \pref{masslessnoise} in the case of a massless scalar.

We now assemble these mode-specific results into a statement about the entire density matrix for the field. To this end we use the fact that each mode is independent (for free fields on a de Sitter background) and so the total density matrix can be written
\be
 \Xi := \prod_k \otimes \rho_k
\ee
where $\rho_k :=   \Psi_k \,   \Psi^*_k$ is the density matrix under discussion up to this point, which we now know satisfies its individual FP equation, eq.~\pref{noisyFPk}. This implies the diagonal matrix elements of $\Xi$ satisfy
\bea \label{XiFPeq}
 \frac{\partial}{\partial t} \, \langle \chi_{k1}, \chi_{k2}, \dots | \Xi | \chi_{k1}, \chi_{k2}, \cdots \rangle &=& \sum_q \frac{\partial}{\partial t} \, \langle \chi_q | \rho_q | \chi_{q} \rangle \left[ \prod_{k \ne q} \otimes \langle \chi_k | \rho_k | \ \chi_k \rangle \right] \\
 &=& \sum_q \cN_q(t) \, \frac{\partial^2}{\partial \chi_q \partial \chi_q^*} \, \langle \chi_q | \rho_q | \chi_{q} \rangle \left[ \prod_{k \ne q} \otimes \langle \chi_k | \rho_k | \ \chi_k \rangle \right] \nn\\
 &=& \sum_q \cN_q(t) \, \frac{\partial^2}{\partial \chi_q \partial \chi_q^*} \, \langle \chi_{k1}, \chi_{k2}, \dots | \Xi | \chi_{k1}, \chi_{k2}, \cdots \rangle  \,,\nn
\eea

We next wish to rewrite this equation in position space, to make contact with the traditional formulation. To do so we use box normalization while still within the Schr\"odinger picture, so that the mode expansion reads
\be\label{deffourg}
 \chi(x) = \frac{1}{L^{3/2}} \sum_k \chi_k \, e^{ikx}
 \quad \hbox{and} \quad
 \chi_k = \frac{1}{L^{3/2}} \int \exd^3x \; \chi(x) \,e^{-ikx} \,,
\ee
where $L^3$ is the co-moving volume of the box (which drops out of all physical quantities).

In the spirit of tracking only super-Hubble modes we follow that part of the field that is approximately constant only over a particular Hubble volume, $\Omega$, and not globally:
\bea\label{defoom}
 \chi_{\Omega}(r) &:=& \frac{1}{\Omega} \int_{\Omega} \exd^3x \, \sqrt{-g} \; \chi(r+x) \nn\\
 &=& \frac{1}{L^{3/2}} \sum_k  \chi_k  \, f_\Omega(k) \, e^{ikr}
 \,,
\eea
where $r$ represents the coarse-grained coordinate label that identifies which Hubble volume we follow. The quantity
\be
 f_\Omega(k) := \frac{1}{\Omega} \int_\Omega \exd^3x \sqrt{-g} \; e^{ikx} \,,
\ee
is a masking function that vanishes for $k/a \gg H$ and is close to unity when $k/a \ll H$, though we show below that our results do not depend on its detailed form and apply equally well for any other masking function that shares these two limits.

This coarse-graining can be regarded as a marginalization over sub-Hubble modes, and so changes the diagonal density matrix elements to
\be\label{cpmarg}
\cP\,=\, \left( \prod_k \, \frac{a^3 (\talpha + \talpha^*)}{\pi S_\Omega(k) } \right) \exp \left[ - a^3 \sum_k \left( \frac{\talpha + \talpha^*}{S_\Omega(k)} \right) \chi^*_k \, \chi_k \right] \,,
\ee
with $S_\Omega = \left| f_\Omega \right|^2$ being the window function\footnote{Although it might seem odd to find the window function in the denominator, this expresses the omission of those modes for which $|S_\Omega|\to 0$, since their variance goes to zero.} described above that accepts modes with $k/a \ll H$ but rejects those with $k/a \gg H$. This window function, $S_\Omega(k)$, changes the Fokker-Planck equation, eq.~\pref{XiFPeq}, satisfied by $\cP$ in two separate ways. First, the Jacobian
\be\label{jaco}
 \frac{\partial}{\partial \chi_k} = \frac{f_\Omega(k) \, e^{ikr} }{L^{3/2}}  \; \frac{\partial}{\partial \chi_\Omega(r)} \,,
\ee
implies we have
\bea \label{Nsum}
 \sum_q \cN_q(t) \, \frac{\partial^2}{\partial \chi_q \partial \chi_q^*} &=& \frac{1}{L^3} \sum_q  |f_\Omega(q)|^2 \cN_q(t) \, \frac{\partial^2}{\partial \chi_\Omega^2} \nn\\
 &=&\int \frac{\exd^3q}{(2\pi)^3} \; S_\Omega(q) \cN_q(t) \, \frac{\partial^2}{\partial \chi_\Omega^2} \,,
\eea
where we suppress the dependence on $\vec r$. The second effect of the window function, $f_\Omega(k)$, arises because it is a function of the scale factor, $a$, and so its presence introduces a new source of time-dependence into $\cP$ {\em in addition} to the time-evolution predicted by the Schr\"odinger evolution of the field state. For the gaussian wave-functionals of interest, explicit calculation shows this new contribution to $\partial_t \cP$ remains proportional to $\partial^2 \cP/\partial \chi_k \partial \chi_k^*$, and so represents a second contribution to the noise coefficient which must be summed with eq.~\pref{Nsum}.

Combining these leads to the following coarse-grained position-space Fokker-Planck equation,
\be \label{posspaceFP}
 \frac{\partial}{\partial t} \, \cP[ \chi_\Omega ] = \cN_\Omega \, \frac{\partial^2 \, \cP[ \chi_\Omega ]}{\partial \chi_\Omega^2} \,,
\ee
for the super-Hubble diagonal matrix elements, $P[\chi_\Omega] := \langle {\chi}_{\Omega} | \Xi | \chi_\Omega \rangle$,
which describes the evolution of the probability density $P[\chi_\Omega ]$, for finding a particular coarse-grained scalar field, $\chi_\Omega$. The coefficient $\cN_\Omega$ evaluates to
\bea \label{XiFPeq2}
 \cN_\Omega &=& \partial_t \int \frac{\exd^3 q}{(2\pi)^3} \; S_\Omega(q) \langle \chi_q^* \, \chi_q \rangle 
 = \frac{H}{4\pi^2} \int \exd q \; q^2 ( a \partial_a)  \Big[ S_\Omega(q) \, |w_q(a)|^2 \Bigr]   \nn\\
 &=& \frac{H}{4\pi^2} \int \exd q \; q^2 \left\{ \frac{1}{a^3} (a \partial_a)  \Big[ S_\Omega(q) \,a^3 |w_q(a)|^2 \Bigr] - 3 S_\Omega(q) \, |w_q(a)|^2 \right\} \nn\\
 &=& \frac{H}{4\pi^2} \int \exd q \; q^2 \left\{ \frac{1}{a^3} (- q \partial_q)  \Big[ S_\Omega(q) \,a^3 |w_q(a)|^2 \Bigr] - 3 S_\Omega(q) \, |w_q(a)|^2 \right\} \nn\\
 &=& -\frac{H}{4\pi^2} \int \exd q \; \partial_q  \Bigl\{ q^3 \, S_\Omega(q) \, |w_q(a)|^2  \Bigr\} \nn\\
 &=& \frac{H}{4\pi^2} \lim_{q \to 0}  q^3 \, |w_q(a)|^2  = \frac{H}{4\pi^2} \lim_{q \to 0}  \left( \frac{q^3}{a^3} \right) \, \frac{1}{\omega_q + \omega_q^*} \,,
\eea
where the first line uses $\partial_t = aH \partial_a$ and integrates over $2\pi$ solid angle (which, since we use complex modes satisfying $\chi_q^* = \chi_{-q}$, is required to avoid double-counting). The third line then uses that $S_\Omega(q)$ and $a^3 |w_q(a)|^2$ depend on $q$ and $a$ only through the combination $x = q/aH$, and the final limit uses the properties that the window function satisfies $S_\Omega(q \to \infty) = 0$ and $S_\Omega(q \to 0) = 1$. This way of writing things emphasizes that $\cN_\Omega$ is independent of any other precise details of the window function.

Specializing eq.~\pref{XiFPeq2} to a massless scalar field, using eq.~\pref{masslessw}, then gives
\be \label{XiFPeq3a}
 \cN_\Omega \simeq \frac{H}{4\pi^2} \lim_{q \to 0}  \frac{H^2}{2} \left[ 1 + \left( \frac{q}{aH} \right)^2 \right] = \frac{H^3}{8\pi^2}  \,.
\ee
Using this result eq. (\ref{posspaceFP}) reduces to
\be \label{posspaceFPc}
 \frac{\partial}{\partial t} \, \cP[ \chi_\Omega ] = \left( \frac{H^3}{8\pi^2}  \right) \frac{\partial^2 \, \cP[ \chi_\Omega ]}{\partial \chi_\Omega^2} \,,
\ee
which is the standard result for the noise term in absence of interactions. The noise term is seen here to be sensitive only to the zero-momentum limit of the mode functions and independent of the precise details of how the window function turns on near $q = aH$.

\subsection*{Modifications arising for sub-luminal sound speeds}

As a fairly trivial application of the tools developed to this point we explore how the stochastic noise, eq.~\pref{posspaceFPc}, changes if the scalar field under discussion should have a sub-luminal (but constant) sound speed $c_s\le 1$. Models of this type have been studied both to explain inflation models as well as dark energy.

Suppose the scalar Lagrangian is again eq.~(\ref{Ldef}), but with generalized dispersion relation
\be
  M_k^2\,=\,\frac{c_s^2 k^2}{a^2}\,,
\ee
where $c_s \le 1$ is a constant sound speed. For simplicity we also neglect the mass, $m=0$. In this case all calculations of the previous subsections can be repeated almost verbatim, with the result that the function $\omega$ of eq.~(\ref{masslessomega}) is now given by
\be\label{newom}
  \omega(a)_{m\to 0}\,=\,
    \frac{c_s^3\,k^3}{a \left[ c_s^2\, k^2 + (aH)^2 \right]} + \frac{iH\,c_s^2\, k^2}{c_s^2\,k^2 + (aH)^2}
  \,,
\ee
and so eq.~\pref{XiFPeq3a} gives
\be
 {\cal N}_\Omega\,=\,\frac{H^3}{8 \,\pi^2\,c_s^3}\,,
\ee
showing that a smaller sound speed {\em enhances} the noise coefficient. Why the speed of sound does so can be seen from the integral expression for $\cN_\Omega$ evaluated with a step-function for $S_\Omega(q)$,
\bea
 \label{XiFPeq3}
 \cN_\Omega &\simeq& \frac{1}{4\pi^2} \left\{ i \int_0^{\frac{aH}{c_s}} \frac{\exd q\, q^2}{a^3} \left( \frac{\talpha - \talpha^*}{\talpha + \talpha^*} \right) + \left[ \left( \frac{q^2}{a^3} \right) \frac{a H^2 }{\talpha + \talpha^*} \right]_{q = {aH}/{c_s}} \right\} \,.
\eea
The speed of sound, $c_s$, strongly modifies this formula because it changes the scale of horizon exit to $c_s k\,=\,a H$, since it is the crossing of the sound horizon that is relevant to the freezing of the modes.

\subsection{Getting the drift}\label{sec-drift}

We next move a step further along the slow-roll expansion to allow a nontrivial classical background about which the quantum fluctuations occur. We here show how these add the standard drift contribution to the probability distribution's evolution equation, (\ref{posspaceFPc}).

To this end we add to the scalar action a potential term,
\be
 L = \int \exd^3 x \, a^3({t}) \, \left[\frac12\,
 \dot \chi^2 -\frac{1}{2\,a^2(t)} (\nabla \chi)^2
 -V(\chi)\right] \,. \label{gena1}
\ee
The conjugate momentum is as before, $\Pi\,=\,a^3(t) \,\dot{\chi}$, while the corresponding Hamiltonian density is
\be
 {\cal H}\,=\,  \frac{\Pi^2}{2\,a^3(t)}
 +\frac{a(t)}{2}\,\left( \nabla \chi\right)^2+a^3(t)\,
 V(\chi)\,.
\ee

We now drop the assumption that the classical background be time-independent, and instead consider fluctuations about a homogeneous classical background, $\chi_b(t)$, which we assume satisfies the classical field equations in the slow-roll regime. That is, we assume the potential in the region of interest satisfies the slow-roll conditions
\be \label{slowrollcond}
 \frac12 \left[ \frac{(V')^2}{V} \right] \ll 3H^2
 \quad \hbox{and} \quad
  V''  \ll 3H^2 \,,
\ee
which would become the usual slow-roll conditions if we were also to assume the Hubble scale satisfies $3H^2 = V/M_p^2$ (where $M_p = (8\pi G)^{-1/2}$ is the Planck mass). We avoid this additional assumption here so as to be able to include spectator scalar fields in addition to the inflaton itself. These slow-roll conditions allow the neglect of $\ddot \chi_b$ relative to $H \dot \chi_b$ and so imply $\chi_b$ is related to $V$ by the usual expression
\be \label{slrc}
 \,\dot{ \chi}_b\simeq -\frac{V'}{3 H}\,.
\ee

The full scalar field can be written
\be\label{decob}
  \chi(t,\vec x)\,=\,\chi_b(t)+ \hat{\chi}(t,\vec x)\,,
\ee
where the quantum perturbation, $\hat \chi$, also includes a homogeneous, $k=0$, component
\be \label{debc0}
 \chi_0(t) = \chi_b(t) + \hat\chi_0(t)\,,
\ee
because it is also allowed to fluctuate around the classical background. Physically, fluctuations in this constant mode can be regarded as describing the ensemble of different Hubble volumes, given that one focusses on expectations and correlations for observables that all refer only to a single Hubble volume.

Using eq.~(\ref{decob}) in the Hamiltonian density allows its Schr\"odinger representation to be expressed as  \cite{Boyanovsky:1993xf}
\be
 {\cal H}\,=\, -\frac{1}{2 \,a^3(t)}\,\frac{\delta^2}{\delta\,\hat \chi^2}
 +\frac{a(t)}{2}\,\left( \nabla \hat  \chi\right)^2+a^3(t)\left[ V( \chi_b)+V'(\chi_b)\,\hat \chi+\frac12\,V''(\chi_b)
 \,\hat \chi^2+\dots
 \right]
\ee
where to leading order in slow-roll we may neglect more than two derivatives of the potential.\footnote{A potential term in $\cH$ linear in $\hat \Pi$ can be removed by a canonical transformation, leading to a linear term in the gaussian wave-functional considered below.} As before this breaks up into independent evolution for each Fourier mode, and the only sector that cares about the time-dependent background is the zero-mode, whose Hamiltonian becomes (up to an additive constant)
\be \label{hamifo}
 {\cal H}_0 =
  -\frac{1}{2 \,a^3(t)}\,\frac{\partial^2}{\partial\, \chi_0^2}
 +a^3(t) \left[ V'(\chi_b)\,  \chi_0 + \frac{1}{2}\,V''(\chi_b)
 \, \chi_0^2 +\cdots \right] \,.
\ee
Because of the force term proportional to $V'$ we generalize the gaussian ansatz for the wave-function of this mode to
\be
 \Psi[\chi_0] = N(t)\,\exp{\left\{ - a^3(t) \left[ \frac12 \, \,\omega(t) \, \chi_0^2 + b(t)\, \chi_0
   \right] \right\} }\label{anspsi} \,,
\ee
where $\omega(t)$, $b(t)$ and $N(t)$ are complex functions, although $\chi_0$ is real.

\subsubsection*{Time evolution}

The time-evolution of the coefficients $N_k$, $\omega_k$, $b_0$ is determined as before, by plugging the ansatz (\ref{anspsi}) into the Schr\"odinger equation. This leads to the same evolution equation as before for the coefficient $\omega$,
\be
 \dot\omega +3\, H\,\omega +i\,\omega^2-i V''
  \simeq 0 \,,
 \label{eqome}
\ee
and
\be
 \dot b  + 3\,H\,b +i\,\omega\,b-\,i\, V'
 \simeq 0 \label{eqofb}\,,
\ee
where we drop contributions suppressed by higher derivatives of $V$ or higher powers of $ V'$ or $ V''$, as appropriate for leading order in slow-roll.

Because the equation for $\omega_k$ is the same as before, so is the solution (and in particular the solution with the correct asymptotics in the past satisfies $\omega \to 0$ as $k \to 0$). The new information comes from the time-evolution of the real and imaginary parts of $b$, whose time-independent solutions are very simple,
\be
 b = \frac{iV'}{3 H} \, \,.
\ee
Although derived for de Sitter space, this solution also provides the leading contribution to the function $b$ for near-de Sitter geometries within the slow-roll regime.

\subsubsection*{Fokker-Planck equation (single mode)}

As in the previous subsections, our real interest is in deriving the evolution equation for the diagonal density matrix $\langle \chi_0 |\tilde{\rho}| \chi_0 \rangle$. The Schr\"odinger equation implies the required evolution law for this quantity is
\bea\label{schme}
 \partial_t \, \langle \chi_0 | \rho | \chi_0 \rangle &=& \left[
 \frac{i}{2a^3} \left( \frac{\partial^2}{\partial \xi_0^2}
 -\frac{\partial^2}{\partial \chi_0^2} \right) \langle \xi_0 |\rho| \chi_0 \rangle  \right]_{\xi_0=\chi_0} \, \,.
\eea

It remains to evaluate the derivatives in the right hand side. Writing
\be
 P_0[\chi_0] := \langle \chi_0 |\rho| \chi_0 \rangle = |N|^2 \, \exp{ \left\{ - a^3 \left[ \frac12 \, (\omega +\omega^*) \,\chi_0^2 + \left( b +b^* \right) \,
   \chi_0 \right] \right\} }
\ee
a direct calculation (see Appendix \ref{app:gaussian}) shows that
\bea \label{longformula}
 \left[
 \frac{i}{2a^3} \left( \frac{\partial^2}{\partial \xi_0^2}
 -\frac{\partial^2}{\partial \chi_0^2} \right) \langle \xi_0 |\rho| \chi_0 \rangle  \right]_{\xi_0=\chi_0}
 &=& \frac{i}{2} \left\{ - (\omega - \omega^*) + a^3 \Bigl[ \Bigl( \omega^2 - (\omega^*)^2 \Bigr) \chi_0^2 \right. \\
 && \qquad \qquad \left. + 2 \Bigl( \omega b - \omega^* b^* \Bigr) \chi_0 + \Bigl( b^2 - (b^*)^2 \Bigr) \Bigr] \right\}P_0 \,,\nn
\eea
while
\be
 \frac{\partial P_0}{\partial \chi_0} = - a^3 \Bigl[ (\omega + \omega^* ) \chi_0 + (b + b^*) \Bigr] P_0 \,,
\ee
and so on for $\partial^2 P_0/\partial \chi_0^2$.

In these expressions we are to use $b = iV'/3H$, with $V' = V'(\chi_b)$ evaluated at the classical background, $\chi_b$. However, we wish to express the evolution of $P_0(\chi_0)$ in terms of $V'(\chi_0)$, and so must expand
\be
 V'(\chi_b) = V'(\chi_0) - \hat \chi_0 \, V''(\chi_0) \,,
\ee
which amounts to shifting $b \to b + c \; \hat \chi_0$ with $b = iV'(\chi_0)/3H$ implying $c = -i V''(\chi_0)/3H$. After performing this shift, repeating the steps of Appendix \ref{app:gaussian} allow the right-hand side of eq.~\pref{longformula} to be exchanged for $P_0$ and its derivatives. Unlike the case of pure noise the presence of $b$ and $c$ introduce terms involving one and no derivatives of $P_0$ into the expression for $\partial_t P_0$. Because $b$ and $c$ are pure imaginary these terms simplify considerably --- {\em c.f.} eq.~\pref{ABCwhenBCimaginary} --- to give
\bea
 \Bigl( \partial_t P_0 \Bigr)_{\rm drift} &=& \frac{i}{2a^3} \left[ \Bigl( -2 a^3 b \Bigr) \left( \frac{\partial P_0}{\partial \chi_0} \right) + \Bigl( 2 a^3 c \Bigr) P_0 \right] \nn\\
 &=&  \frac{V'(\chi_0)}{3H} \left( \frac{\partial P_0}{\partial \chi_0} \right) + \left( \frac{V''(\chi_0)}{3H}  \right) P_0 \nn\\
 &=& \frac{1}{3H} \frac{\partial}{\partial \chi_0} \Bigl[ V'(\chi_0) \, P_0(\chi_0) \Bigr],
\eea
as well as a modification to the second-derivative (noise) term of the form
\bea
 \Bigl( \partial_t P_0 \Bigr)_{\rm noise} &=& \frac{i}{a^3} \left( \frac{\omega - \omega^*}{\omega + \omega^*} + \frac{2c}{\omega+\omega^*} \right) \left( \frac{ \partial^2 P_0}{\partial \chi_0^2} \right) \nn\\
 &=& i |w(a)|^2 \left( \omega - \omega^* - \frac{2i V''(\chi_0)}{3H} \right) \left( \frac{ \partial^2 P_0}{\partial \chi_0^2} \right) \,.
\eea

\subsubsection*{Complete Fokker-Planck equation}

We now collect the contribution of different momentum modes, as before, to get the total drift for the coarse-grained field $\chi_\Omega$. Defining, as before, the coarse-grained probability distribution, $${\cal P}\,\equiv\,\prod_k\,P_k\,,$$ and using the results of the previous subsection to express the noise term, we see that $\cP$ satisfies the evolution equation
\be\label{cofor}
 \partial_t \cP = \cN_\Omega \, \frac{\partial^2 \cP}{\partial  \chi_\Omega^2} + \cD_\Omega \, \frac{\partial}{\partial \,  \chi_\Omega} \,\left( \frac{\partial V (\chi_\Omega)}{\partial  \chi_\Omega} \, {\cal P} \right)
 \,,
\ee
where the drift coefficient is given by the usual expression
\be \label{driftresult}
 \cD_{\Omega} =
 \frac{1}{3 H} \,,
\ee
while the noise coefficient is modified to
\be
 \cN_\Omega = \cN_\Omega^{(0)} + \cN_\Omega^{(1)} \,,
\ee
where
\be \label{N0}
 \cN_\Omega^{(0)} = \frac{H}{4\pi^2} \; \lim_{k \to 0} \, k^3 |w|^2 \,,
\ee
is the result obtained earlier --- {\em c.f.} eq.~\pref{XiFPeq2} --- and
\be \label{N1}
 \cN_\Omega^{(1)} := -  \frac{2 V''(\chi_0)}{3H} \int \exd^3 q \; S_\Omega(q) |w|^2
 = - \frac{ V''(\chi_0)}{6\pi^2H}  \int \exd q \; q^2 S_\Omega(q) |w|^2 \,.
\ee
Eqs.~(\ref{cofor}) through \pref{N1} reproduce the standard Starobinsky result including the contributions of drift,\footnote{Since our arguments rely explicitly on the Gaussian wave-functional, our derivation as presented here is insufficient to derive the Fokker-Planck equation far from the Gaussian regime, where it is also believed to hold.} although most applications (see for instance \cite{Finelli}) neglect $\cN^{(1)}$ and the contributions of particle masses to $\cN^{(0)}$ (more about which below).

\subsection{Mass-dependent noise}

\begin{figure}[t,b,h]
  \centering
  \includegraphics[width=0.7\textwidth]{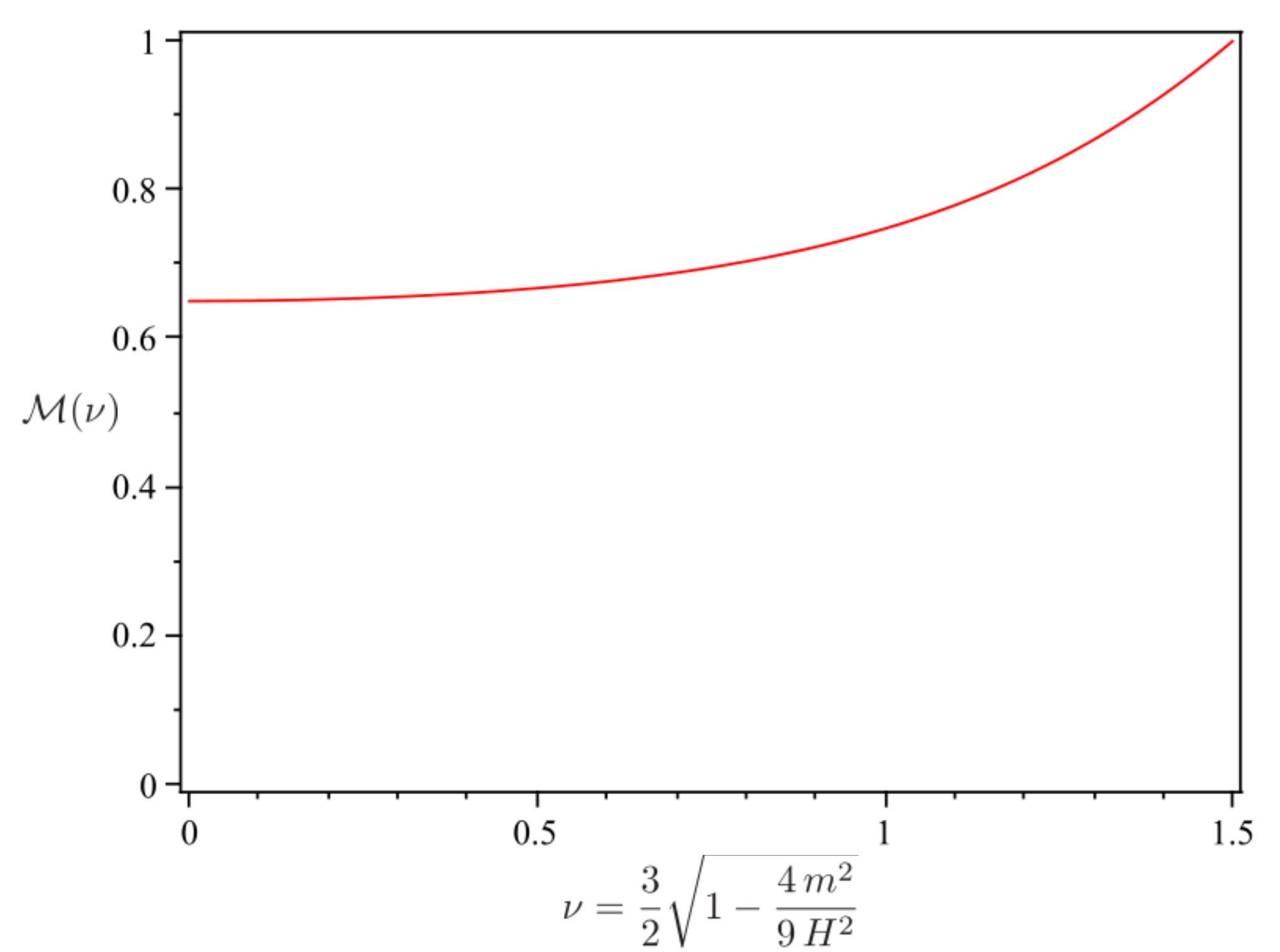}
  \caption{Plot of $\cM(\nu)$ for values of $\nu$ from 0 to 3/2 (corresponding to values of $m$ from $3H/2$ to 0, respectively).}
\label{fig:Mplot}
\end{figure}

It is common in the literature to drop the term $\cN_\Omega^{(1)}$ and work with the massless limit for the noise, which turns out not to be a bad approximation for small masses (see figure \ref{fig:Mplot}). However the term $\cN_\Omega^{(1)}$ plays an important conceptual role in demonstrating this, as we now pause to explore. Along the way we give compact expressions for the noise as a function of mass. To this end we specialize to the case where $V''(\chi_0) = m^2$ is independent of $\chi_0$.

A naive determination of the mass-dependence of the noise term would simply use the massive mode-functions, eq.~\pref{massivew}, in expression \pref{N0}. However, this leads to a puzzle due to the divergent small-$k$ limit of the massive mode functions. For $m^2 \ll H^2$ the relevant small-$k$ limit of the massive mode functions is
\bea
 \lim_{k\to 0} k^3 |w|^2 &=& \lim_{k\to 0} \frac{\pi k^3}{4a^3H} \left( \frac{1}{\sin(\pi\nu) \Gamma(1-\nu)} \right)^2 \left( \frac{2aH}{k} \right)^{2\nu} \nn\\
 &=& \frac{H^2}{2} \lim_{k\to 0}\left\{ 1 - \frac{2m^2}{3H^2} \left[ \log \left( \frac{2aH}{k} \right) + \cC \right] + \cO \left( \frac{m^4}{H^4} \right) \right\} \,,
\eea
where $\cC$ is a calculable constant and $\nu$ is as defined below eq.~\pref{massivebessel}. This shows a strong sensitivity to the ordering of the limits $k \to 0$ and $m^2 \to 0$, due to the IR divergence in the expansion of $\cN_\Omega^{(0)}$ in powers of $m^2/H^2$:
\be
 \left( \cN_\Omega^{(0)} \right)_{IR} = \frac{H^3}{8\pi^2} \left( \frac{2m^2}{3H^2} \right) \lim_{k \to 0} \log \left( \frac{k}{aH} \right) \,.
\ee

It is the term $\cN_\Omega^{(1)}$ that cures this divergence, leaving a nonsingular expression for $\cN_\Omega$ for small masses. To see this we use a simple step window-function, $S_\Omega(q) = \Theta(q - aH)$, in which case
\be
 \cN_\Omega^{(1)} = \frac{H}{4\pi^2} \left( \frac{2m^2}{3H^2} \right) \int_{0}^{aH} \exd q \, q^2   |w|^2 \,.
\ee
To leading order in $m^2/H^2$ we may evaluate this using the massless mode function, to get
\be
 \cN_\Omega^{(1)} \simeq \frac{H}{8\pi^2} \left( \frac{2m^2}{3H^2} \right) \int_{0}^{aH} \frac{ \exd q }{ q} \left[ 1 + \left( \frac{q}{aH} \right)^2 \right] \,,
\ee
whose divergent lower limit
\be
 \left( \cN_\Omega^{(1)} \right)_{IR} = - \frac{H^3}{8\pi^2} \left( \frac{2m^2}{3H^2} \right) \lim_{k \to 0} \log \left( \frac{k}{aH} \right) \,,
\ee
precisely cancels the divergence in $\cN_\Omega^{(0)}$ when this is expanded in powers of $m^2/H^2$.

Combining terms the full result for the noise can be compactly written
\be \label{conciseN}
 \cN_\Omega = \frac{H^3}{8\,\pi^2} \; {\cal M}(\nu)
\ee
where
\be \label{conciseM}
 {\cal M}(\nu) :=  \frac{|\Gamma(\nu)|^2\,(3+2\nu)}{3\,\pi\,2^{2-2 \nu}} + \frac{(9-4\nu^2)  }{12}\,K(\nu) \,,
\ee
and the integral
\be
 K(\nu) :=
 \pi\int_{0}^{1} \exd x \,  x^2  \left[ \left| H_\nu^{(1)}\left( x \right) \right|^2 - \left| \frac{1}{\sin(\pi\nu) \Gamma(1-\nu)} \right|^2 \left( \frac{x}{2} \right)^{-2\nu} \right] \,.
\ee
$K(\nu)$ is defined so as to have a smooth limit as $m \to 0$ --- and normalized such that $K(\nu) \to 1$ as $\nu \to 3/2$ --- making it straightforward to integrate numerically (as is done for the plot in Figure \ref{fig:Mplot}). As the figure shows, the dependence of the noise on particle mass is quite weak throughout the entire range $0 \le m \le \frac32 \, H$.

\section{Interactions and decoherence}
\label{sec:Interactions}

We next generalize the previous section's discussion to include interactions between the short- and long-wavelength modes. We do so within the open EFT formalism by setting up and solving the appropriate Lindblad equation for inflationary cosmology. We find the main effect of interactions is to drive off-diagonal elements of the density matrix rapidly to zero, thereby decohering the initially quantum system into a classical stochastic one whose effects are what are observed in CMB measurements.

\subsection{Generalization to FRW Geometries}

We start by restating the coarse-grained Lindblad equation of \S\ref{sec:OpenEFT}, which reads
\bea \label{dmeqbody2}
    &&\frac{\partial \rho_\ssA}{\partial t}
    = aH \frac{\partial \rho_\ssA}{\partial a}
   =
    i  
  \; \Bigl[ \rho_\ssA , \cA_j
    \Bigr] \, \langle \cB^j
    \rangle_\ssB
    -  \frac12 
    \; \cW^{jk} \Bigl[
    \cA_j \cA_k \rho_\ssA + \rho_\ssA \cA_j \cA_k
    - 2 \cA_k \rho_\ssA \cA_j \Bigr]  \,,
\eea
to second order in $\tau_c/\tau_p$. For curved-space applications we notice that there can be metric dependence hidden in this expression, such as if --- for local interactions --- the implied sum in contractions like $\cA_i \cB^i$ include an integration over space. For interactions that transform as Lorentz scalars the sum then contains a factor of the 3-metric's
volume element, as in
\be \label{Vdef2}
    \cA_i(t) \cB^i(t) = a^3(t) \int \exd^3x \; \cA_a(x,t) \, \cB^a(x,t) \,,
\ee
where $a,b$ are ordinary indices that run over a finite range. If $\cA_a$ and $\cB^a$ should instead be tensors there are also powers of $a$ coming from contraction of spatial indices: {\em e.g.} $A_\mu B^\mu = - A_t B_t + \delta^{bc} A_b B_c/a^2$. But the rapid growth of $a$ inexorably red-shifts away these alternative sources of $a$, leading again to the $a$-dependence of eq.~\pref{Vdef2}.

Eq.~\pref{dmeqbody2} is the equation whose solution we seek. To this end we first dispense with the first term on its right-hand side. Recall that we work in the interaction picture for which operators evolve under the influence of the terms $\cH_\ssA + \cH_\ssB$ while states involve under the influence of $V$. Since the term linear in $V$ in eq.~\pref{dmeqbody2} can be regarded as a mean-field modification of $\cH_\ssA$,
\be
 \ol \cH_\ssA := \cH_\ssA +
    \cA_i \, \langle \cB^i \rangle_\ssB \,,
\ee
it is useful to regard it as a part of $\cH_\ssA$ for the purposes of defining the interaction representation. Once this is done the linear term drops from the right-hand side of eq.~\pref{dmeqbody2}, and we track only the second-order term in the evolution of the redefined $\rho_\ssA$.

\subsection{Solutions}

After the redefinition of interaction representation just described, the time-evolution of $\rho_\ssA$ to be solved becomes
\bea \label{dmeqbody3}
    &&\frac{\partial \rho_\ssA}{\partial t}
    = aH \frac{\partial \rho_\ssA}{\partial a}
    =  -
    \frac12\; \cW^{jk} \Bigl[ \cA_j \cA_k \rho_\ssA + \rho_\ssA \cA_j \cA_k
    - 2 \cA_k \rho_\ssA \cA_j \Bigr] \,,
\eea
where the time-dependence of the operators in $\cA_i$ are now to be evaluated using $\ol \cH_\ssA$ rather than $\cH_\ssA$.

Although red-shifting quickly forces all spatial derivatives to zero in $\cH$, in general the operators $\cA_i$ can involve both the fields and their canonical momenta:
\be \label{Achipi}
 \cA_i = \cA_i (\chi, \Pi) \,.
\ee
However because the fields evolve with time according to $\cH_\ssA$ (or $\ol\cH_\ssA$) they experience the expansion of the background geometry and in particular, for weakly interacting fields, their modes become squeezed in the standard way \cite{squeezed}. But this squeezing drives the canonical momenta to zero, so after a few Hubble times eq.~\pref{Achipi} is instead well-approximated by
\be \label{Achi}
 \cA_i \simeq \cA_i (\chi, 0) \,.
\ee
This is an enormous simplification because it implies that all of the operators $\cA_i(x,t)$ commute with one another at equal times, because they are all diagonal in the field basis,
\be \label{alphadef}
 \cA_i | \chi \rangle = \alpha_i(\chi) | \chi \rangle \,.
\ee

Because of this eq.~\pref{dmeqbody3} is most easily integrated by taking its matrix elements in the field basis, for which the operators $\cA_i(x,t)$ are diagonal. Denoting in this basis
\be
    \rho_t[\chi,\tilde\chi] =  {}_t\langle \chi| \rho_\ssA|
    \tilde\chi  \rangle_t \,,
\ee
and using the interaction-picture evolution $\partial_t |\chi
\rangle_t = -i\ol\cH_\ssA |\chi\rangle_t$, we find
\bea \label{drhodteqn}
    \frac{\partial\rho_t}{\partial t}[\chi,\tilde\chi]
    &=& (\partial_t \langle \chi |) \rho_\ssA |\tilde\chi \rangle
    + \langle \chi | \partial_t \rho_\ssA |\tilde\chi \rangle
    + \langle \chi | \rho_\ssA (\partial_t |\tilde\chi
    \rangle)\nonumber\\
    &=& \langle \chi | \partial_t \rho_\ssA |\tilde\chi \rangle
    + i \langle \chi | [\rho_\ssA, \ol\cH_\ssA] |\tilde\chi \rangle \nn\\
    &=& \left( \frac{\partial\rho_t}{\partial t} \right)_0 + \langle \chi | \partial_t \rho_\ssA |\tilde\chi \rangle \,.
\eea
We roll the terms involving $\ol\cH_\ssA$ into the first term because we are most interested in the {\em difference} in evolution due to the terms in eq.~\pref{dmeqbody3}. In particular, because the $\ol\cH_\ssA$ terms describe Hamiltonian evolution they cannot generate a mixed state from one which is initially pure, unlike the terms coming from eq.~\pref{dmeqbody3}, whose decoherence effects we wish to follow.

Using eq.~\pref{dmeqbody3} to evaluate the matrix elements of $\partial_t \rho_\ssA$ we find
\be \label{dmeqscalarcosmologyme}
    \frac{\partial \rho_t}{\partial t} =
    \left( \frac{\partial\rho_t}{\partial t} \right)_0
     - \rho_t 
     \Bigl[
    \alpha_i - \tilde\alpha_i
    \Bigr]\Bigl[ \alpha_j - \tilde\alpha_j \Bigr] \, \cW^{ij}(t) \,,
\ee
where $\alpha_i = \alpha_i(\chi)$ and $\tilde \alpha_i = \alpha( \tilde \chi)$ are the eigenvalues defined in eq.~\pref{alphadef}. The general solution is easily given, in the form
\be
    \rho_t[\chi,\tilde\chi] = \rho^{0}_t[\chi,\tilde\chi]
    \, e^{-\Gamma} \,,
\ee
where $\rho^0_t$ is the pure-state result obtained using only $\ol\cH_\ssA$ and
\be \label{GaussianForm}
    \Gamma = \int_{t_0}^t \exd t'  
     \Bigl[
    \alpha_i - \tilde\alpha_i \Bigr]\Bigl[\alpha_j
    - \tilde\alpha_j \Bigr]
    \cW^{ij}(t') \,.
\ee

This shows how terms involving the fluctuations $\cW^{ij}$ cause the reduced density matrix to take the form of a classical Gaussian distribution in the $\chi$ basis, whose time-dependent width is controlled by the local autocorrelation function $\cW^{ij}$. Provided this width shrinks in time, at late times the system evolves towards a diagonal density matrix (in the $|\chi \rangle$ basis), with diagonal probabilities that are set purely by the $\ol\cH_\ssA$ evolution,
\be
    P_t[\chi] \equiv \rho_t[\chi,\chi]
    = \rho^0_{t}[\chi,\chi] = \Bigl| \Psi^0_{t}[\chi]
    \Bigr|^2 \,,
\ee
where $\Psi^0_{t}[\alpha]$ is the wave-functional for the initial
pure state, whose evolution is as given in the free-field evolution of \S\ref{sec:FreeFields}.

We see the effect of the quadratic terms in eq.~\pref{dmeqbody3} is to decohere the initial state into the classical stochastic ensemble for the field variables, $\{ \chi \}$, just as is usually assumed when analyzing CMB observables in terms of inflationary fluctuations at horizon re-entry. What ultimately makes the field basis special in this regard is the squeezing of super-Hubble modes, which acts to drive to zero all canonical momenta in the interactions contained in $V$.  The environment on whose ignorance this decoherence relies consists of all of those sub-Hubble modes of the field within the many Hubble volumes with which the super-Hubble modes of interest overlap.

\subsection{Decoherence Rates}

Attractive as the above picture is, it leaves two questions open. One of these asks what the time-dependence is of the gaussian in $\Gamma$. Implicit in the decoherence story is the assumption that its variance shrinks with time, rather than grows. Even should this be true a second question arises: do 50--60 $e$-folding furnish enough time to decohere initially quantum fluctuations between horizon exit and subsequent re-entry? Does a sufficiently strong decoherence rate provide any constraint on the interactions of a putative inflaton?

This section addresses these issues, arguing that the gaussian variance indeed shrinks (rather than grows) over Hubble time-scales under very broad assumptions about the structure of the interactions. It further argues that this shrinking is very fast, being essentially complete in just a few Hubble times even for the gravitational strength interactions that any primordial fluctuation source must have.

The argument proceeds largely on dimensional grounds. For simplicity we assume $V$ has the form of eq.~\pref{Vdef}, where $\cB^i$
is a local operator having engineering dimension (mass)${}^d$,
\be \label{Vdeflocal}
    V(t) = \cA_i(t) \, \cB^i(t) = \int \exd^3 x \,a^3(t) \cA_a(x,t) \, \cB^a(x,t)\,,
\ee
and so the evolution equation, eq.~\pref{dmeqbody3}, becomes
\bea \label{dmeqbody3local}
    \frac{\partial \rho_\ssA}{\partial t}
    &=&  -  \frac{1}{2} \int \exd^3x \exd^3x'\;  a^6(t) \,
     \cW^{ab}(x,x',t) \nn\\
     && \qquad\qquad \times \Bigl[ \cA_a(x) \cA_b(x') \rho_\ssA + \rho_\ssA \cA_a(x) \cA_b(x')
    - 2 \cA_b(x') \rho_\ssA \cA_a(x) \Bigr] \,.
\eea
where
\be \label{shortdistcorrslocal}
    \langle \delta {\cal B}^a(x,t) \, \delta {\cal B}^b(x',t')
    \rangle_\ssB = \cW^{ab}(x,x',t) \,
    \delta(t-t') \,.
\ee
In eq.~\pref{dmeqbody3local} it is the delta-correlation of eq.~\pref{shortdistcorrslocal} that ensures that operators like $\cA_a(x,t)$ are all evaluated at the same time, but even this time-dependence becomes trivial after a few Hubble times in the extra-Hubble regime due to the squeezing of the states, eq.~\pref{Achi}.

The locality assumption is a natural one, and makes the time-dependence easier to track. But because our conclusions are essentially dimensional we believe they apply more generally than this. On dimensional grounds the factor $\cA_i$ then has dimension (mass)$^{4-d}$. With these assumptions the function $\cW^{ab}(x,x',t)$ defined by eq.~\pref{shortdistcorrslocal} then has dimension (mass)${}^{2d-1}$. As an example, a gravitational-strength interaction between a field $\chi$ and sector $B$ could have the form $\chi \, \cB /M_p$, for some operator $\cB$. If $\chi$ is a canonically normalized field having mass dimension one then $\cB$ has dimension (mass)${}^4$, as appropriate for a stress-energy density (say), and $\cW(x,x',t)$ would have dimension (mass)${}^7$.

Suppose now that the physics of sector $B$ is characterized by a single mass scale, $\Lambda(t)$, which can be slowly evolving with time as the universe expands. This would be true in particular for a trace over sub-Hubble modes in the usual Bunch-Davies style vacuum. For instance, for massless modes $\Lambda(t)$ might be given by the Hubble scale itself, $H$. Alternatively it might be described by the temperature, $T(t)$, if $B$ were described by a simple thermal state. For simplicity suppose also that $\cW^{ab}$ is a function only of the proper distance, $\sigma(x,x')= a(t) s(x,x')$, between the spatial points $x$ and $x'$ (as is often the case, such as for homogeneous systems). On dimensional grounds we then have
\be \label{ScaleEstimate}
    \cW^{ab}(x, x', t) = c^{ab}[a(t) s(x, x') \Lambda(t)] \,\Lambda^{2d-1}(t) \,,
\ee
where $c^{ab}$ are calculable dimensionless real functions. For a broad class of stable environments we can take $c^{ab} \ge 0$ to be positive definite.

Using these expressions we find that eq.~\pref{dmeqbody3local} becomes
\bea \label{dmeqscalar}
    \frac{\partial \rho_\ssA}{\partial t} &=&  - \frac12 \int \exd^3x \exd^3x'\;  a^6(t) \, c^{ab}[\sigma(x, x') \Lambda] \,\Lambda^{2d-1}(t)\,
\; \Bigl[
    \cA_a(x), \Bigl[ \cA_b(x') , \rho_\ssA \Bigr] \Bigr] \nn\\
   &=&  - \frac12 \int \exd^3x \, \exd^3 u \; a^3(t) \, c^{ab}[u] \,\Lambda^{2d-4}(t)\,
\; \Bigl[
    \cA_a(x), \Bigl[ \cA_b(x,u) , \rho_\ssA \Bigr] \Bigr]  \,,
\eea
where the second line changes integration variables from $x'$ to $u = \sigma(x,x') \Lambda$. With these choices the gaussian argument, $\Gamma$, of eq.~\pref{GaussianForm} is then
\be \label{GaussianFormScalar}
    \Gamma \sim \int \exd^3x \, \exd^3u\;  c^{ab}(u) \int_{t_0}^t \exd t' \; a^3(t') \, \Lambda^{2d-4}(t')
   \Bigl[ \alpha_a(x) - \tilde\alpha_a(x) \Bigr] \Bigl[ \alpha_b(x,u) - \tilde\alpha_b(x,u) \Bigr]
    \,.
\ee
The combination $e^{-\Gamma}$ has the form of a Gaussian functional of $\alpha_i(x) - \tilde\alpha_i(x)$, which when evaluated for a configuration that takes the value $\alpha_0$ extending over a physical volume of order
$a^3(t)\,L^3$ is
\be
    e^{-\Gamma}
    \sim \exp\left[-\, \frac{L^{3}
    (\alpha_0 - \tilde{\alpha}_0)^2}{\sigma^2(t)}
    \right]  \,,
\ee
with a width, $\sigma(t)$, that evolves in time according to
\be
    \frac{1}{\sigma^2} \sim
    \int^t_{t_0} dt' \; \Lambda^{2d-4}(t') \, a^{3}(t')
    = \int^a_{a_0} \frac{\hat{a}^2 d\hat{a}}{H(\hat{a})} \;
    \Lambda^{2d-4}(\hat{a})
    \,.
\ee

In the present instance, for effectively massless modes ($m \ll H$) coherent over a Hubble volume we expect $\Lambda \sim H$, so $\Lambda$ evolves with time as does $H$ (and so is nearly time-independent during an inflationary epoch). In this case the Gaussian density matrix acquires the time evolution
\be
    \frac{1}{\sigma^2} \sim
   \int^a_{a_0} \frac{\hat{a}^2 d\hat{a}}{H(\hat{a})} \;
    \Lambda^{2d-4}(\hat{a}) \sim \left( \frac{\Lambda_0^{2d-4}}{H} \right) a^3
    \,,
\ee
which clearly grows exponentially like the volume, $a^3 \sim e^{3Ht}$, during a near-de Sitter inflationary epoch. This shows that the width of the gaussian distribution shrinks exponentially rapidly in cosmic time during inflation, regardless of the precise dimension of the interactions appearing in $V$. What is important is that the scale $\Lambda$ be approximately independent of time, as it would be if it were set by $H$ (or a more microscopic mass scale).

Knowledge of the time-dependence of $\sigma$ allows an estimate of whether sufficient time passes after horizon exit to decohere the modes of interest for CMB observations. To this end we take for the massless super-Hubble sector-$A$ field the dimensional estimate $\alpha - \tilde \alpha \simeq H^{4-d}$ and $L^3 \sim H^{-3}$ to find
\be
    \frac{L^3 (\alpha - \tilde \alpha)^2}{\sigma^{2}}
    \sim \left(
    \frac{\Lambda_0}{H}
    \right)^{2d-4} a^3\,.
\ee
Adequate decoherence requires $L^3 (\alpha - \tilde{\alpha})^2 / \sigma^2 \gg 1$, and so we see that it is very quickly more than adequate on very general grounds, with only a logarithmic sensitivity to the ratio $\Lambda_0/H$.

\section{Summary and other possible applications}
\label{sec:concl}

Our goal in this paper was to apply the systematic tools of effective field theory to super-Hubble physics in accelerating cosmologies. It has been a long-standing puzzle how to integrate long-distance cosmology into an EFT framework. We believe that this search has been difficult because what is usually sought is an effective {\em lagrangian} rather than just an effective field theory description.

\subsection{Summary of the argument}

We argue here (and in \cite{Companion}) that in general an effective lagrangian description is inappropriate when EFT techniques are applied to open systems (for which the degrees of freedom integrated out are able to exchange information with those being explicitly followed). The effective description of particles moving through a fluid (such as photons through macroscopic media or neutrinos moving through the Sun \cite{BM,nufluct}) provide examples of this type, and a lagrangian description can fail because such systems can decohere, with pure states evolving to mixed states.

For open systems there is an analog for the simplifications that usually arise in the low-energy limit of traditional effective field theories. The corresponding simplifications arise instead in the long-time limit, for which interaction time-scales, $\tau_p$, are long compared with the correlation times, $\tau_c$, of the interactions with the environment. In this case time evolution can be computed perturbatively in powers of $\tau_c/\tau_p$, leading to a Lindblad equation for the evolution of a reduced density matrix. This equation can also usually be integrated to infer reliably the long-time evolution over scales $t \gg \tau_p$.

We argue here that the Lindblad equation is the correct language for describing the EFT of super-Hubble modes, since modes continually move from inside to outside the Hubble scale. In this case the environment consists of sub-Hubble modes, whose correlation time is the Hubble time, $\tau_c \sim H^{-1}$. The hierarchy of scale on which simplifications rely is obtained by focussing on evolution that takes place on much longer times, $H \tau_p \gg 1$.

We set up the Lindblad formalism for super-Hubble modes and use it to evolve the reduced density matrix for these modes. This evolution is most easily tracked in a basis of field eigenstates because the state-squeezing of super-Hubble modes drives their canonical momenta to zero, thereby ensuring that the field basis always diagonalizes the super-Hubble/sub-Hubble interactions. In this basis we find that off-diagonal matrix elements of the reduced density matrix are robustly driven to zero over a few Hubble times, ensuring that initially quantum fluctuations decohere while frozen outside the Hubble scale, and that they re-enter as classical in the field basis at later epochs (as is assumed when analyzing their implications for CMB observations).

For weak interactions the diagonal elements of the reduced density matrix are well-described by their free evolution in the presence of the gravitational background. In this limit the Lindblad equation reduces to the usual Liouville equation, and we show that this implies the diagonal matrix elements evolve according to the standard Fokker-Planck equation of Starobinsky's stochastic formulation of inflation. Taken together, these results for diagonal and off-diagonal elements provide a systematic derivation of stochastic inflation as the leading part of the super-Hubble EFT (a result which will not surprise cosmologists, least of all Starobinsky).

There are several novelties in our derivation. In particular: ($i$) we show that decoherence occurs very quickly, even for interactions of gravitational strength; ($ii$) we provide the framework to derive the {\em deviations} from the leading stochastic behaviour (which is often where EFT methods really come into their own); and ($iii$) we show (as do standard treatments) why the stochastic noise is generated as successive modes leave the Hubble scale and freeze, but do so in a way that allows one to see why the result is independent of the details of the window function used to distinguish super- from sub-Hubble wavelengths.

\subsection{Future directions: IR resummations, secular behaviour and black holes}

We believe open EFTs can be very useful in repurposing the powerful tools of EFTs to cases where information gets exchanged between a system and its environment. In particular it potentially provides a more robust statement of the precise domain of validity of standard calculations.

\subsubsection*{Late-time resummations}

Because the stochastic picture governs the long-distance and late-time behaviour, it provides the natural language in which to cast the resummation of IR divergences in de Sitter (and near-de Sitter) space \cite{TsamisWoodard}, and the associated late-time secular evolution. We intend to explore this connection in more detail in future work. 

In particular, an important part of our derivation of the Fokker-Planck equation is the Gaussian ansatz for the quantum state of the quantum field of interest, and important use was also made (as is also true for standard derivations) of free-field correlation functions when deriving the noise term in the Langevin equation for long-wavelength modes. Because of this our derivation is unable to directly verify the validity of Starobinsky's use \cite{CMBthn} of this equation in the late-time limit to yield very non-gaussian probability distributions. Such a derivation would require a late-time resummation of accumulating IR effects --- akin to the dynamical renormalization group used in ref.~\cite{deSRG} --- and we hope in future to better understand the role of Gaussianity in the late-time validity of the stochastic framework.

\subsubsection*{Potential relevance to black holes}

Black holes provide another example where the Open EFT language is likely to be useful for understanding gravitational systems. In this case the system of interest is outside the black hole, and the environment is inside. And again information is exchanged between these two systems.
What is useful about recognizing these as open EFTs (rather than just garden-variety low-energy EFTs) is that this identifies that there are more conditions underlying the validity of approximations beyond the standard assumptions of low energy and small curvatures. In particular, in addition to these the validity of the Lindblad equation relies on the two conditions given in \S\ref{subsec:openhier}.

Of particular interest in the case of black holes is condition 2, which states that the environment ({\em i.e.}~the black hole) remain unperturbed by the evolution of the external system. Unlike the usual low-energy, small-curvature conditions, this assumption breaks down near the Page time, once appreciable information begins to escape the black hole, even in regions far from the curvature singularity.

We believe that the proper EFT formulation for the standard calculations of quantum black-hole behaviour is likely to rely on the Lindblad equation, and so its breakdown near the Page time is likely to be a useful point of departure for understanding the counter-intuitive puzzles to do with black-hole information loss. At the very least it is likely to provide an understanding of why standard calculations start to break down, even in regimes where curvatures remain small.

Now that we have a new hammer, let's go find all those nails...

\section*{Acknowledgements}

We thank Robert Brandenberger, Subodh Patil, Laurence Perreault Levasseur, Fernando Quevedo, Albert Roura, Andrew Tolley, Mike Trott, Vincent Vennin, Mark Wyman and Itay Yavin for helpful conversations about effective theories, decoherence and cosmology. Robert Brandenberger in particular has long advocated the point of view we adopt (and place on a precise footing) in this paper. C.B. thanks the Abdus Salam International Centre for Theoretical Physics for its hospitality at various points during the completion of this work, and we all especially thank the Research in Teams program at the Banff International Research Station for providing such pleasant environs and helping us to secure the undivided time that allowed us (at last!) to bring this long-standing project to completion. R.~H.~was supported in part by DOE grant DE-FG03-91-ER40682 and by a grant from the John Templeton Foundation, while the research of C.B.~and M.W.~is partially supported by grants from N.S.E.R.C.~(Canada) and Perimeter Institute for Theoretical Physics. Research at Perimeter Institute is supported in part by the Government of Canada through Industry Canada, and by the Province of Ontario through the Ministry of Research and Information (MRI).  G.T. thanks
STFC for financial support through the grant ST/H005498/1.

\appendix

\section{Solving for time dependence}
\label{app:solving}

This appendix solves the time-evolution equations to obtain explicit expressions for the quantity $\talpha$. For convenience we first repeat the equations to be solved, eqs.~\pref{talphaeq} and \pref{tNeq} of the main text:
\be \label{app:talphaeq}
 \dot \talpha + 3 H \talpha = -i  \talpha^2 +i M^2 \,,
\ee
and
\be \label{app:tNeq}
 \dot N = -i\talpha \, N\,,
\ee
where
\be
 M^2 = \frac{k^2}{a^2} + m^2 \,,
\ee
inherits its $t$-dependence from $a$.

\subsection*{The case of constant $M$}

We first integrate these equations in the case where $a$ (and so also $M$) is constant.

\subsubsubsection{Static solution}

\bigskip\noindent
We define the static solution to be the one for which $\talpha$ is constant in time. In this case inspection of eq.~\pref{app:talphaeq} reveals the static solution to be
\be
 \talpha(t) = \pm M  \,,
\ee
which is real in the usual harmonic oscillator case, for which $M^2 > 0$. In this section we also entertain the case of the inverted oscillator, $M^2 < 0$, for which the static solution is imaginary. This time-dependent solution is the energy eigenstate (and when $M^2 > 0$ is in particular the ground state when $\talpha = + M > 0$).

\subsubsubsection{Time-dependent solutions}

\bigskip\noindent
We can also integrate the more general case, corresponding to situations where the initial state is not prepared in the vacuum. When $M$ is time-independent eq.~\pref{talphaeq} can be solved using the elementary integral
\be
  \int \frac{\exd x}{b^2 - x^2} = \frac{1}{b} \, \tanh^{-1} \left( \frac{x}{b} \right) \,,
\ee
leading to the solution (for approximately constant $M$),
\bea
 \talpha(t) &=& \cM \tanh \Bigl[ i M (t - t_0) + \beta \Bigr] \nn\\
 &=&  M \left\{ \frac{ \tanh[ i M(t - t_0) ] + \tanh \beta}{1 + \tanh[i M(t - t_0)] \tanh \beta} \right\} \\
 &=& \frac{ \talpha_0 + i M \tan [ M (t - t_0) ]}{1 + i (\talpha_0/M) \tan[M(t-t_0)]} \,,\nn
\eea
where $M \tanh \beta := \talpha(t = t_0) := \talpha_0$ (which requires $|\talpha_0| \le |M|$ if $\beta$ is real). The real and imaginary parts of this expression are
\bea
 \hbox{Re}\; \talpha(s) &=& \talpha_0 \left\{ \frac{1 + \tan^2[M(t-t_0)]}{1 + (\talpha_0/M)^2 \tan^2[M(t-t_0)]} \right\} \nn\\
 &=& \frac{\talpha_0}{\cos^2[M(t-t_0)] + (\talpha_0/M)^2 \sin^2[M(t-t_0)]} \\
 \hbox{and} \quad \hbox{Im}\; \talpha(s) &=& M \left\{ \frac{[1 - (\talpha_0/M)^2]\tan[ M(t-t_0)]}{1 + (\talpha_0/M)^2 \tan^2[M(t-t_0)]} \right\} \,, \nn
\eea
where it is assumed that $\talpha_0$ and $M$ are real. Notice the real part remains positive if it is initially positive, and that $\talpha$ is time-independent if $\talpha_0 = M$ (which is the case of an adiabatic vacuum). For later use what is useful is the ratio of imaginary and real parts, which has the simple oscillatory form
\be
 \frac{\talpha - \talpha^*}{\talpha + \talpha^*} = \frac{i}{2} \left( \frac{M}{\talpha_0} - \frac{\talpha_0}{M} \right) \sin \left[ 2 M (t - t_0) \right] \,.
\ee

\subsubsubsection{The inverted oscillator}

\bigskip\noindent
The above way of writing the solution is most useful when $M$ is real, but although $M^2$ must be real we are also interested in the case where it is negative: $M^2 = - \mu^2 < 0$, and so $M = i \mu$. In this case the above formulae are more usefully written
\be
 \talpha(t) = \frac{ \talpha_0 - i \mu \tanh [ \mu (t - t_0) ]}{1 + i (\talpha_0/\mu) \tanh[\mu(t-t_0)]} \,,
\ee
whose real and imaginary parts are
\bea
 \hbox{Re}\; \talpha(s) &=& \talpha_0 \left\{ \frac{1 - \tanh^2[\mu(t-t_0)]}{1 + (\talpha_0/\mu)^2 \tanh^2[\mu(t-t_0)]} \right\} \nn\\
 &=& \frac{\talpha_0}{\cosh^2[\mu(t-t_0)] + (\talpha_0/\mu)^2 \sinh^2[\mu(t-t_0)]} \nn\\
 \hbox{and} \quad \hbox{Im}\; \talpha(t) &=& -\mu \left\{ \frac{[1 + (\talpha_0/\mu)^2]\tanh[ \mu(t-t_0)]}{1 + (\talpha_0/\mu)^2 \tanh^2[\mu(t-t_0)]} \right\} \,,
\eea
where it is again assumed that $\talpha_0$ is real. These are no longer oscillatory (though the real part remains positive). In the late-time limit, $\mu (t - t_0) \to \infty$, they asymptote to
\be
 \hbox{Re}\; \talpha(t) \to \left[ \frac{4 \talpha_0}{1 + (\talpha_0/\mu)^2} \right] \; e^{-2\mu(t-t_0)} \to 0 \,,
\ee
and
\be
 \hbox{Im}\; \talpha(t) \to -\mu \Bigl[ 1 + \cO(e^{-\mu(t-t_0)}) \Bigr] \,,
\ee
corresponding to the static solution, $\talpha = -i \mu$. (No static solution is possible if $\talpha_0$ is real.) The transition from real to imaginary $\talpha$ corresponds to the change-over from a wavefunction localized about zero to a delocalized state whose wavefunction is pure phase.

In this asymptotic solution we have
\be
 \frac{\talpha - \talpha^*}{\talpha + \talpha^*} =  - \frac{i}{4} \left( \frac{\talpha_0}{\mu} + \frac{\mu}{\talpha_0} \right) e^{+ 2 \mu (t - t_0)}  \,,
\ee
which grows without bound, starting from a place that is invariant under $\talpha_0 \to \mu^2/\talpha_0$.

\subsection*{Time-varying $M$}

We now turn to the general case where $a$ varies in time, so that the quantity $M$ becomes a function of time. We start by integrating \pref{app:talphaeq}.

\subsubsubsection{Static solutions}

\bigskip\noindent
In this situation the static solutions ({\em i.e.} those satisfying $\dot \talpha = 0$) satisfy
\be
 i  \talpha^2 + 3H \talpha - i M^2 = 0 \,,
\ee
which has as roots
\be
 \talpha_\pm = \frac12 \Bigl( 3iH \pm \sqrt{4M^2 - 9H^2} \Bigr) \,,
\ee
and so $\talpha_\pm \to \pm M$ for $M \gg H$, and $\talpha_+ \to 3iH$ and $\talpha_- \to \cO(iM^2/H)$ for $M \ll H$. Notice these are both imaginary and their imaginary parts have the same sign (provided $M^2 > 0$).

\subsubsubsection{Time-dependent solutions}

\bigskip\noindent
To solve more generally change variables from $t$ to $a$ using $\dot\talpha = a H \talpha'$, where $y' := \exd y/\exd a$, and define the new variable $v = -i \talpha/aH$ to put \pref{app:talphaeq} into Riccati form
\be
 v' = v^2 - \frac{4v}{a} + \frac{M^2}{a^2 H^2} \,.
\ee
This can be made into a linear second-order equation through the change of variables $v = - w'/w$, giving
\be \label{app:wlineq}
 a^2 w'' + 4a \,w' + \left( \frac{M^2}{H^2} \right) w = a^2 w'' + 4a\, w' + \left[ \left( \frac{k}{aH} \right)^2 + \frac{m^2}{H^2}  \right] w = 0 \,.
\ee

In fact, this is not just any old linear equation, this is the Klein-Gordan equation on de Sitter space, $\exd s^2 = - \exd t^2 + e^{2Ht} \, \exd x^2$, after making the mode-function substitution,
\be
 \varphi(x,t) = \frac{w(t)}{L^{3/2}} \, e^{i k x} \,,
\ee
where we normalize to a box with co-moving size $L$. Then
\bea
 0 = -\frac{1}{\sqrt{-g}} \partial_\mu \left( \sqrt{-g} \; g^{\mu\nu} \partial_\nu \varphi \right) + m^2 \varphi &=& \frac{1}{L^{3/2}}  \Bigl( \ddot w + 3H \dot w + M^2 w \Bigr) e^{ikx} \nn\\
 &=& \frac{1}{L^{3/2}} \Bigl[ H^2 \left( a^2 w'' + 4 a w' \right) + M^2 w \Bigr] e^{ikx}  \,.
\eea
Because of this, spacetime integration of $\sqrt{-g} \Bigl(\varphi^* \Box \varphi - \varphi \,\Box \varphi^* \Bigr) = 0$ shows that the Wronskian,
\be
 \cW(w,w) := \frac{1}{L^{3}} \int_{\Sigma_t} \exd^3x \, \sqrt{-g} \; n^{\mu} \Bigl( w^* \partial_\mu w - w \partial_\mu w^* \Bigr) = a^3(t)  \Bigl( w^* \dot w - w \dot w^* \Bigr) \,,
\ee
is independent of the particular time-slice, $\Sigma_t$ (with unit normal $n^\mu = \delta^\mu_t$), along which the integral is performed. Consequently its value can be chosen to satisfy the mode-normalization condition, $\cW(w,w) = i$, at all times.

Eq.~\pref{app:wlineq} is recognized as a Bessel equation after the change of variable,
\be
 w(x) := x^{3/2} y(x) \,,
\ee
with $x := k/aH$, since then
\be
 x^2 \frac{\exd^2 y}{\exd x^2} + x \frac{\exd y}{\exd x} + \left( x^2 - \nu^2 \right) y = 0 \,,
\ee
with
\be
 \nu^2 = \frac94 - \frac{m^2}{H^2} \,.
\ee

The boundary condition for the remote past is to take the quasi-static solution there, so $\talpha \simeq M \to (k/a)$ as $a \to 0$. Consequently $v = -ik/a^2H$ for small $a$ and so
\be
 w \to w_0 \, \exp \left( - \frac{ik}{aH} \right) = w_0 \, e^{-ix}
 \qquad \hbox{(as $a \to 0$ or $x \to \infty$)} \,.
\ee
This picks out the Hankel functions, $y = H_\nu (x) := H^{(2)}_\nu(x)$, as the appropriate solution, since
\be
 H_\nu^{(2)} (x) \simeq \sqrt{ \frac{2}{\pi x}} \; e^{-i \left( x - \frac{\pi \nu}{2} - \frac\pi 4 \right)} \,,
\ee
in the large-$x$ limit. We find $w(x) \propto x^{3/2} H_\nu(x)$ and so
\be
 \talpha(a) = iH \left[ \frac32 + x \, \frac{\exd}{\exd x} \ln H_\nu \right]_{x = k/aH} \,.
\ee

At late times $a \to \infty$ and so $x \to 0$, the Hankel function has the limiting form
\bea
 H_\nu(x) &\propto& J_{-\nu}(x) - e^{i\pi \nu} J_\nu(x) \nn\\
 &\to& \,  \left( \frac{x}{2} \right)^{-\nu} \left[\, \sum_{n=0}^\infty \frac{(-)^n}{n! \Gamma(n-\nu+1)} \left( \frac{x}{2} \right)^{2n} \right] \\
  &&\qquad\qquad - e^{i\pi\nu} \left( \frac{x}{2} \right)^{\nu} \left[\, \sum_{n=0}^\infty \frac{(-)^n}{n! \Gamma(n+\nu+1)} \left( \frac{x}{2} \right)^{2n} \right]  \,,\nn
\eea
and so
\be \label{lateform}
 \talpha \to iH \left\{ \frac32 - \nu \left[ \left( 1 + \frac{x^2}{2\nu(1-\nu)} + \cdots \right) -2 i \sin(\pi
 \nu) \frac{\Gamma(-\nu)}{\Gamma(\nu)} \left( \frac{x}{2} \right)^{2\nu} \Bigl(1 + \cdots \Bigr) \right] \right\}_{x=k/aH}\,,
\ee
as $a \to \infty$ or $x \to 0$.

The real part of the bracket comes from the subdominant terms in each Bessel function in the small-$x$ limit, while the imaginary part comes from the `other' Bessel function, leading to the following small-$x$ limits:
\be
 \hbox{Re} \, \talpha \simeq 2H \sin(\pi \nu) \frac{\Gamma(-\nu)}{\Gamma(\nu)} \left( \frac{x}{2} \right)^{2\nu} + \cdots \,,
\ee
and
\be
 \hbox{Im} \, \talpha \simeq H \left[ \left( \frac32 - \nu \right) - \frac{x^2}{2(1- \nu)} + \cdots \right] \,.
\ee

In the special case of negligible mass, $m^2 \to 0$, we have $\nu \to \frac32$ and the the above asymptotic form simplifies to
\bea \label{lateform32}
 \talpha &\to& iH \left\{ \frac32 - \frac32 \left[ \left( 1 - \frac23 \, x^2 + \cdots \right) + \frac{2i}{3} \, x^3 \Bigl(1 + \cdots \Bigr) \right] \right\}_{x=k/aH} \nn\\
 &=& iH \Bigl[ \Bigl( x^2 + \cdots \Bigr) - i \Bigl( x^3 + \cdots \Bigr) \Bigr]_{x=k/aH} \,,
\eea
which can be compared directly with the relevant Hankel function, $H_{3/2}(x)$, which has a simple closed-form expression,
\be
 H_{3/2}(x) \propto \frac{1}{\sqrt x} \left( 1 - \frac{i}{x} \right) \, e^{-ix} \,,
\ee
and so $w(x)$ reduces to the very simple expression
\be
 w(x) \propto x^{3/2} H_{3/2}(x) \propto ( x - i ) \, e^{-ix} \,,
\ee
which gives
\be
 \talpha = iH \left( \frac{x^2}{1 + i x} \right)_{x=k/aH}
 = \left[ \frac{H x^2 (i + x)}{1 + x^2} \right]_{x=k/aH} \,.
\ee
In particular, in the massless case we have the result
\be \label{app:masslessnoise}
 i \left( \frac{\talpha - \talpha^*}{\talpha + \talpha^*}\right)_{x=k/aH} = - \left(\frac{aH}{k} \right) \,,
\ee
for all $x$.

\section{Gaussian facts}
\label{app:gaussian}

Consider the general gaussian wavefunction
\be\label{ansapb}
 \Psi[\phi] = N \, \exp \Bigl[ - \cA \, \phi^* \phi + \cB \phi + \overline \cB \phi^* \Bigr] \,,
\ee
which satisfies
\bea
 \langle \Phi \rangle &=& N^2 \int_{-\infty}^\infty \exd \phi^* \exd \phi \; \phi \, \exp \Bigl[ - (\cA+\cA^*) \, \phi^* \phi + (\cB + \ol\cB^*) \phi + (\cB^* + \overline \cB) \phi^* \Bigr] \nn\\
 &=& N^2 \int_{-\infty}^\infty \exd \hat\phi^* \exd \hat\phi \; (u + \hat\phi) \, \exp \Bigl[ - (\cA+\cA^*) \, (u+\hat\phi)^* (u+\hat \phi) \nn\\
 && \qquad\qquad\qquad\qquad\qquad\qquad + (\cB + \ol\cB^*) (u+ \hat\phi) + (\cB^*+\overline \cB) (u+ \hat\phi)^* \Bigr] \nn\\
 &=& u + N^2 \int_{-\infty}^\infty \exd \hat\phi^* \exd \hat\phi \; \hat\phi \, \exp \Bigl\{ - (\cA+\cA^*) \, \hat\phi^* \hat\phi + \Bigl[ - (\cA+\cA^*) u^* + \cB + \ol\cB^*\Bigr] \hat\phi \\
 && \quad + \Bigl[ - (\cA+\cA^*) u + \cB^* + \ol\cB \Bigr] \hat\phi^* + \Bigl[ - (\cA + \cA^*) u^*u + (\cB + \ol\cB^*) u + (\cB^*+\overline \cB) u^* \Bigr] \Bigr\} \nn\\
 &=& u  \nn
\eea
where the last equality holds because of the antisymmetry of the integrand under $\hat \phi \to - \hat \phi$, if we choose
\be
 u = \frac{\cB^* + \ol\cB}{\cA + \cA^*} \,.
\ee

The corresponding density matrix is
\bea
 R(\psi,\phi) &:=& \langle \psi | \rho | \phi \rangle = \Psi[\psi] \; \Psi^*[\phi] \nn\\
 &=& N^2 \, \exp \Bigl[ - \cA \, \psi^* \psi - \cA^* \phi^* \phi + \cB \psi + \overline \cB \psi^* + \cB^* \phi + \ol\cB^* \phi^* \Bigr] \,.
\eea
The derivatives are
\be
 \frac{\partial R}{\partial \psi^*} = \Bigl( - \cA \psi + \ol\cB \Bigr) R \quad \hbox{and} \quad  \frac{\partial R}{\partial \phi^*} = \Bigl( - \cA^* \phi + \ol\cB^* \Bigr) R \,,
\ee
and so
\bea
 \frac{\partial^2 R}{\partial \psi \partial \psi^*} &=& \left[ - \cA + \Bigl( - \cA \psi + \ol\cB \Bigr)  \Bigl( - \cA \psi^*  + \cB \Bigr)  \right] R \nn\\
 &=& \Bigl[ \cA^2 \psi^* \psi - \cA \cB \psi - \cA \ol\cB \psi^*  + \Bigl( \cB \ol\cB - \cA \Bigr) \Bigr] R \,,
\eea
and
\bea
 \frac{\partial^2 R}{\partial \phi \partial \phi^*} &=& \left[ - \cA^* + \Bigl( - \cA^* \phi + \ol\cB^* \Bigr)  \Bigl( - \cA^* \phi^*  + \cB^* \Bigr)  \right] R \nn\\
 &=& \Bigl[ (\cA^*)^2 \phi^* \phi - \cA^* \cB^* \phi - \cA^* \ol\cB^* \phi^*  + \Bigl( \cB^* \ol\cB^* - \cA^* \Bigr) \Bigr] R \,.
\eea
Using these definitions we have
\bea \label{derivdiff}
 \left( \frac{\partial^2 R}{\partial \psi \partial \psi^*} - \frac{\partial^2 R}{\partial \phi \partial \phi^*} \right)_{\psi = \phi} &=& \Bigl\{ \Bigl[ \left[\cA^2 - (\cA^*)^2 \right] \phi^* \phi  - (\cA - \cA^*) \Bigr] - (\cA \cB - \cA^* \cB^*) \phi \nn\\
 && \qquad\qquad  - (\cA \ol\cB - \cA^* \ol\cB^*) \phi^* + (\cB \ol\cB - \cB^* \ol\cB^*) \Bigr\} \cR \,,
\eea
where
\be
 \cR[\phi] := R[\phi, \phi] = \langle \phi | \rho | \phi \rangle = N^2 \exp \Bigl[ - (\cA + \cA^*) \phi^* \phi + (\cB + \cB^*) \phi + (\ol\cB + \ol \cB^*) \phi^* \Bigr] \,.
\ee

We next would like to re-express eq.~\pref{derivdiff} in terms of derivatives of $\cR$, using
\be \label{cRfirstD}
 \frac{\partial \cR}{\partial \phi^*} = \Bigl[ - (\cA+ \cA^*) \phi + (\ol\cB + \ol\cB^*) \Bigr] \cR \,,
\ee
and
\bea
 \frac{\partial^2 \cR}{\partial \phi \partial \phi^*} &=& \Bigl\{ - (\cA+\cA^*) + \Bigl[ - (\cA+\cA^*) \phi + (\ol\cB+ \ol\cB^*)  \Bigr]  \Bigl[ - (\cA+\cA^*) \phi^* + (\cB+ \cB^*)  \Bigr]  \Bigr\} \cR \nn\\
 &=& \Bigl\{ \Bigl[ (\cA+\cA^*)^2 \phi^* \phi - (\cA + \cA^*) \Bigr] - (\cA+\cA^*) (\cB+ \cB^*) \phi \nn\\
  && \qquad\qquad\qquad\qquad - (\cA+\cA^*) (\ol\cB+ \ol\cB^*) \phi^* + ( \cB + \cB^*) (\ol\cB + \ol\cB^*)  \Bigr\} \cR \,.
  \label{secDcR}
\eea
Noticing the identity
\be
 \left[\cA^2 - (\cA^*)^2 \right] \phi^* \phi  - (\cA - \cA^*) = \left( \frac{\cA - \cA^*}{\cA + \cA^*} \right) \Bigl[ (\cA+\cA^*)^2 \phi^* \phi - (\cA + \cA^*) \Bigr]  \,,
\ee
we find
\bea\label{finform}
 \left( \frac{\partial^2 R}{\partial \psi \partial \psi^*} - \frac{\partial^2 R}{\partial \phi \partial \phi^*} \right)_{\psi = \phi} &=& \left( \frac{\cA - \cA^*}{\cA + \cA^*} \right) \frac{\partial^2 \cR}{\partial \phi \partial \phi^*} \\
 &&\qquad + \left(\frac{ \cA^* \cB - \cA \cB^* }{\cA + \cA^*} \right) \frac{\partial \cR}{\partial \phi^*}  + \left(\frac{ \cA^* \ol\cB - \cA \ol\cB^* }{\cA + \cA^*} \right) \frac{\partial \cR}{\partial \phi} \,.\nn
\eea

\subsection*{The real zero mode}

Since the zero mode is real this section briefly records the analog identity for real variables. It also identifies a potentially confusing subtlety that can arise once linear terms are included in the gaussian.

We start with
\be\label{ansapbre}
 \Psi[\chi] = N \, \exp \left[ - \frac12 \, A \, \chi^2 + B \chi \right] \,,
\ee
and density matrix
\be
 R(\xi,\chi) := \langle \xi | \rho | \chi \rangle = \Psi[\xi] \; \Psi^*[\chi]
 = N^2 \, \exp \left[ - \frac12 \, A \, \xi^2 - \frac12 \, A^* \chi^2 + B \xi + B^* \chi  \right] \,.
\ee
The derivatives are
\be
 \frac{\partial R}{\partial \xi} = \Bigl( - A \xi + B \Bigr) R \quad \hbox{and} \quad  \frac{\partial R}{\partial \chi} = \Bigl( - A^* \chi + B^* \Bigr) R \,,
\ee
and
\be
 \frac{\partial^2 R}{\partial \xi^2} = \left[ - A + \Bigl( - A \xi + B \Bigr)^2  \right] R
 = \Bigl[ A^2 \xi^2 - 2 A B \xi  + \Bigl( B^2  - A \Bigr) \Bigr] R \,,
\ee
and
\be
 \frac{\partial^2 R}{\partial \chi^2} = \left[ - A^* + \Bigl( - A^* \chi + B^* \Bigr)^2    \right] R
 = \Bigl[ (A^*)^2 \chi^2 - 2 A^* B^* \chi + \Bigl( (B^*)^2 - A^* \Bigr) \Bigr] R \,.
\ee
Using these definitions we have
\bea \label{derivdiffre}
 \left( \frac{\partial^2 R}{\partial \xi^2} - \frac{\partial^2 R}{\partial \chi^2} \right)_{\xi = \chi} &=& \Bigl\{ \Bigl[ \left[A^2 - (A^*)^2 \right] \chi^2  - (A - A^*) \Bigr] - 2 (A B - A^* B^*) \chi \nn\\
 && \qquad\qquad\qquad\qquad  + \Bigl( B^2 - (B^*)^2 \Bigr) \Bigr\} \cR \,,
\eea
where
\be
 \cR[\chi] := R[\chi, \chi] = \langle \chi | \rho | \chi \rangle = N^2 \exp \Bigl[ - \frac12 (A + A^*) \chi^2 + (B + B^*) \chi  \Bigr] \,.
\ee

As above we wish to re-express eq.~\pref{derivdiffre} in terms of derivatives of $\cR$, using
\be \label{cRfirstDre}
 \frac{\partial \cR}{\partial \chi} = \Bigl[ - (A+ A^*) \chi + (B + B^*) \Bigr] \cR \,,
\ee
and
\bea
 \frac{\partial^2 \cR}{\partial \chi^2} &=& \Bigl\{ - (A+A^*) + \Bigl[ - (A+A^*) \chi + (B+B^*)  \Bigr]^2   \Bigr\} \cR \\
 &=& \Bigl\{ \Bigl[ (A+A^*)^2 \chi^2 - (A + A^*) \Bigr] - 2(A+A^*) (B+ B^*) \chi  + ( B + B^*)^2   \Bigr\} \cR \,.\nn
  \label{secDcRre}
\eea
Combining these we see
\be\label{intmedformre}
 \left( \frac{\partial^2 R}{\partial \xi^2} - \frac{\partial^2 R}{\partial \chi^2} \right)_{\xi = \chi} = \alpha \left( \frac{\partial^2 \cR}{\partial \chi^2} \right) + \beta \left( \frac{\partial \cR}{\partial \chi} \right) + \gamma \, \cR \,,
\ee
where
\be
 \alpha = \frac{A-A^*}{A+A^*} \,, \quad
 \beta = 2 \left( \frac{A^* B  - A B^*}{A + A^*} \right) \quad \hbox{and} \quad \gamma = 0 \,.
\ee

\medskip\noindent{\em A shifted problem}

\medskip\noindent
Now comes the subtle part. Suppose in the above we now shift the coefficient $B \to B + C \chi$ and repeat the above exercise. What is confusing is there are two ways to compute the result that seem naively as if they should be the same, yet which actually differ.

The first way to compute the result is to perform the shift directly in the wave-function,
\be\label{ansapbreshift}
 \Psi[\chi] \to N \, \exp \left[ - \frac12 \, A \, \chi^2 + (B + C \chi) \chi \right] = N \, \exp \left[ - \frac12 \, \hat A \, \chi^2 + B  \chi \right] \,,
\ee
where $\hat A = A - 2C$. Repeating the steps above then shows that eq.~\pref{intmedformre} is true with coefficients given by
\be
 \alpha = \frac{\hat A- \hat A^*}{\hat A+ \hat A^*} \,, \quad
 \beta = 2 \left( \frac{\hat A^* B  - \hat A B^*}{\hat A + \hat A^*} \right) \quad \hbox{and} \quad \gamma = 0 \,.
\ee

A different result is obtained if instead the substitution $B \to B + C \chi$ is made in the expressions for the derivatives of $R$ and $\cR$, such as eqs.~\pref{derivdiffre} and \pref{cRfirstDre}.  In this case under the replacement $B \to B + C \xi$ we instead find
\be
 \frac{\partial R}{\partial \xi} = \Bigl( - A \xi + B \Bigr) R \to \Bigl[ - (A - C) \xi + B \Bigr] R  \,,
\ee
which is `as if' we replace $A$ by $\check A = A - C \ne \hat A$. However for the second derivative we have
\bea
 \frac{\partial^2 R}{\partial \xi^2} = \left[ - A + \Bigl( - A \xi + B \Bigr)^2  \right] R
 &\to& \left\{ - A + \Bigl[ - (A - C) \xi + B \Bigr]^2  \right\} R \\
 &=& \left\{ - \check A + \Bigl[ - \check A \xi + B \Bigr]^2  \right\} R - C R\,,\nn
\eea
and so
\bea \label{derivdiffre2}
 \left( \frac{\partial^2 R}{\partial \xi^2} - \frac{\partial^2 R}{\partial \chi^2} \right)_{\xi = \chi} &=& \Bigl\{ \Bigl[ \left[A^2 - (A^*)^2 \right] \chi^2  - (A - A^*) \Bigr] - 2 (A B - A^* B^*) \chi \nn\\
 && \qquad\qquad\qquad\qquad  + \Bigl( B^2 - (B^*)^2 \Bigr) \Bigr\} \cR \nn\\
 &\to& \Bigl\{ \Bigl[ \left[\check A^2 - ( \check A^*)^2 \right] \chi^2  - (\check A - \check A^*) \Bigr] - 2 (\check A B - \check A^* B^*) \chi \nn\\
 && \qquad\qquad\qquad  + \Bigl( B^2 - (B^*)^2 \Bigr) \Bigr\} \cR - (C - C^*) \cR \,.
\eea

Combining as before we find
\be\label{intmedformre2}
 \left( \frac{\partial^2 R}{\partial \xi^2} - \frac{\partial^2 R}{\partial \chi^2} \right)_{\xi = \chi} = \alpha \left( \frac{\partial^2 \cR}{\partial \chi^2} \right) + \beta \left( \frac{\partial \cR}{\partial \chi} \right) + \gamma \, \cR \,,
\ee
where
\be
 \alpha = \frac{\check A- \check A^*}{\check A+ \check A^*} \,, \quad
 \beta = 2 \left( \frac{\check A^* B  - \check A B^*}{\check A + \check A^*} \right) \quad \hbox{and} \quad \gamma = 2 \left( \frac{AC^* - A^*C}{\check A + \check A^*} \right) \,.
\ee
The important thing is that $\gamma$ is now nonzero. Notice that these expressions simplify in the case of interest $B = - B^*$ and $C = - C^*$, because then $\check A + \check A^* = A + A^*$ and $\check A - \check A^* = A - A^* - 2C$, and so
\be \label{ABCwhenBCimaginary}
 \alpha = \frac{A - A^* - 2C}{A + A^*} \,, \qquad
 \beta = 2 B \qquad \hbox{and} \qquad
 \gamma = -2C \,.
\ee
These are the results used in the main text, with the replacements $A = a^3 \omega$, $B = - a^3 b$ and $C = - a^3 c$.

These methods of calculation differ in their result because differentiations with respect to $\chi$ and $\xi$ were implicitly done {\em with $B$ held fixed}. This means these derivatives differ before and after $B$ is shifted in a $\xi$- or $\chi$-dependent way.


\begin{thebibliography}{99}
\bibitem{EFTs}
   For some reviews aimed at gravity formulated as an EFT see:

 C.~P.~Burgess,
  ``Quantum gravity in everyday life: General relativity as an effective field theory,''
  Living Rev.\ Rel.\  {\bf 7} (2004) 5
  [gr-qc/0311082];

  W.~D.~Goldberger,
  ``Les Houches lectures on effective field theories and gravitational radiation,''
  hep-ph/0701129;

 J.~F.~Donoghue,
  ``The effective field theory treatment of quantum gravity,''
  AIP Conf.\ Proc.\  {\bf 1483} (2012) 73
  [arXiv:1209.3511 [gr-qc]].


\bibitem{TPI}
  C.P. Burgess, J.M. Cline, R. Holman and F. Lemieux,
  ``Are Inflationary Predictions Sensitive to Very High Energy Physics?,''
  JHEP 0302 (2003) 048 [hep-th/0210233];
%
   C.P. Burgess, J.M. Cline and R. Holman,
   ``Effective Field Theories and Inflation,''
  JCAP 0310 (2003) 004 [hep-th/0306079];

 S.~Weinberg,
  ``Effective Field Theory for Inflation,''
  Phys.\ Rev.\ D {\bf 77} (2008) 123541
  [arXiv:0804.4291 [hep-th]];

    G.~Shiu and J.~Xu,
  ``Effective Field Theory and Decoupling in Multi-field Inflation: An Illustrative Case Study,''
  Phys.\ Rev.\ D {\bf 84} (2011) 103509
  [arXiv:1108.0981 [hep-th]];

  A.~Achucarro, J.~-O.~Gong, S.~Hardeman, G.~A.~Palma and S.~P.~Patil,
  ``Effective theories of single field inflation when heavy fields matter,''
  JHEP {\bf 1205} (2012) 066
  [arXiv:1201.6342 [hep-th]];

   A.~Avgoustidis, S.~Cremonini, A.~-C.~Davis, R.~H.~Ribeiro, K.~Turzynski and S.~Watson,
  ``Decoupling Survives Inflation: A Critical Look at Effective Field Theory Violations During Inflation,''
  JCAP {\bf 1206} (2012) 025
  [arXiv:1203.0016 [hep-th]];

  C.~P.~Burgess, M.~W.~Horbatsch and S.~.P.~Patil,
  ``Inflating in a Trough: Single-Field Effective Theory from Multiple-Field Curved Valleys,''
  JHEP {\bf 1301} (2013) 133
  [arXiv:1209.5701 [hep-th]].

\bibitem{InfEFT}
  C.~Cheung, P.~Creminelli, A.~L.~Fitzpatrick, J.~Kaplan and L.~Senatore,
  ``The Effective Field Theory of Inflation,''
  JHEP {\bf 0803} (2008) 014
  [arXiv:0709.0293 [hep-th]];

   L.~Senatore and M.~Zaldarriaga,
  ``The Effective Field Theory of Multifield Inflation,''
  JHEP {\bf 1204} (2012) 024
  [arXiv:1009.2093 [hep-th]];

    N.~Bartolo, M.~Fasiello, S.~Matarrese and A.~Riotto,
  ``Large non-Gaussianities in the Effective Field Theory Approach to Single-Field Inflation: the Bispectrum,''
  JCAP {\bf 1008} (2010) 008
  [arXiv:1004.0893 [astro-ph.CO]];
  %
  ``Large non-Gaussianities in the Effective Field Theory Approach to Single-Field Inflation: the Trispectrum,''
  JCAP {\bf 1009} (2010) 035
  [arXiv:1006.5411 [astro-ph.CO]];

    D.~Lopez Nacir, R.~A.~Porto, L.~Senatore and M.~Zaldarriaga,
  ``Dissipative effects in the Effective Field Theory of Inflation,''
  JHEP {\bf 1201} (2012) 075
  [arXiv:1109.4192 [hep-th]];

   E.~Dimastrogiovanni, M.~Fasiello and A.~J.~Tolley,
  ``Low-Energy Effective Field Theory for Chromo-Natural Inflation,''
  JCAP {\bf 1302} (2013) 046
  [arXiv:1211.1396 [hep-th]].

\bibitem{CMBth1}
 V. F. Mukhanov and G. V. Chibisov, JETP Lett. 33, 532 (1981)
 [Pisma Zh. Eksp. Teor. Fiz. 33, 549 (1981)];

  A. H. Guth and S. Y. Pi, Phys. Rev. Lett. 49, 1110 (1982);

  A. A. Starobinsky, Phys. Lett. B 117, 175 (1982);

  S. W. Hawking, Phys. Lett. B 115, 295 (1982);

  V. N. Lukash, Pisma Zh. Eksp. Teor. Fiz. 31, 631 (1980);
  Sov. Phys. JETP 52, 807 (1980) [Zh. Eksp. Teor. Fiz. 79, (1980)];

  W. Press, Phys. Scr. 21, 702 (1980);

  K. Sato, Mon. Not. Roy. Astron. Soc. 195, 467 (1981).

\bibitem{Sloth}
 M.~S.~Sloth,
  ``On the one loop corrections to inflation
  and the CMB anisotropies,''
  Nucl.\ Phys.\  B {\bf 748}, 149 (2006)
  [arXiv:astro-ph/0604488];
%
  ``On the one loop corrections to inflation. II:
  The consistency relation,''
  Nucl.\ Phys.\  B {\bf 775}, 78 (2007)
  [arXiv:hep-th/0612138].

\bibitem{Bilandzic:2007nb}
  A.~Bilandzic and T.~Prokopec,
  ``Quantum radiative corrections to slow-roll inflation,''
  Phys.\ Rev.\  D {\bf 76}, 103507 (2007)
  [arXiv:0704.1905 [astro-ph]].

\bibitem{vMS}
 M.~van der Meulen and J.~Smit,
  ``Classical approximation to quantum cosmological
  correlations,''
  JCAP {\bf 0711}, 023 (2007)
  [arXiv:0707.0842 [hep-th]].

\bibitem{Petri}
 G.~Petri,
  ``A Diagrammatic Approach to Scalar Field
  Correlators during Inflation,''
  arXiv:0810.3330 [gr-qc].

\bibitem{Lyth:2007jh}
  D.~H.~Lyth,
  ``The curvature perturbation in a box,''
  JCAP {\bf 0712}, 016 (2007)
  [arXiv:0707.0361 [astro-ph]].

\bibitem{Enqvist:2008kt}
  K.~Enqvist, S.~Nurmi, D.~Podolsky and G.~I.~Rigopoulos,
  ``On the divergences of inflationary superhorizon perturbations,''
  JCAP {\bf 0804}, 025 (2008)
  [arXiv:0802.0395 [astro-ph]].

\bibitem{BMPRS}
N.~Bartolo, S.~Matarrese, M.~Pietroni, A.~Riotto and D.~Seery,
  ``On the Physical Significance of Infra-red Corrections
  to Inflationary
  Observables,''
  JCAP {\bf 0801}, 015 (2008)
  [arXiv:0711.4263 [astro-ph]].

\bibitem{RS}
   A.~Riotto and M.~S.~Sloth,
  ``On Resumming Inflationary Perturbations beyond One-loop,''
  JCAP {\bf 0804}, 030 (2008)
  [arXiv:0801.1845 [hep-ph]].

\bibitem{Wbg}
 S.~Weinberg,
  ``Quantum contributions to cosmological correlations,''
  Phys.\ Rev.\  D {\bf 72} (2005) 043514
  [arXiv:hep-th/0506236];
%
  ``Quantum contributions to cosmological correlations.
  II: Can these
  corrections become large?,''
  Phys.\ Rev.\  D {\bf 74} (2006) 023508
  [arXiv:hep-th/0605244];

\bibitem{Adshead:2008gk}
  P.~Adshead, R.~Easther and E.~A.~Lim,
  ``Cosmology With Many Light Scalar Fields: Stochastic Inflation and Loop
  Corrections,''
  Phys.\ Rev.\  D {\bf 79}, 063504 (2009)
  [arXiv:0809.4008 [hep-th]].

\bibitem{UraTan}
  Y.~Urakawa and T.~Tanaka,
  ``Influence on observation from IR divergence
  during inflation -- Multi field
  inflation --,''
  arXiv:0904.4415 [hep-th];
%
  ``No influence on observation from IR divergence during
  inflation I -- single
  field inflation --,''
  arXiv:0902.3209 [hep-th].

\bibitem{Giddings:2010nc}
  S.~B.~Giddings and M.~S.~Sloth,
  ``Semiclassical relations and IR effects in de Sitter and slow-roll space-times,''
  JCAP {\bf 1101} (2011) 023
  [arXiv:1005.1056 [hep-th]].

\bibitem{Byrnes:2010yc}
  C.~T.~Byrnes, M.~Gerstenlauer, A.~Hebecker, S.~Nurmi and G.~Tasinato,
  ``Inflationary Infrared Divergences: Geometry of the Reheating Surface versus $\delta N$ Formalism,''
  JCAP {\bf 1008} (2010) 006
  [arXiv:1005.3307 [hep-th]].


\bibitem{Giddings:2011zd}
  S.~B.~Giddings and M.~S.~Sloth,
  ``Cosmological observables, IR growth of fluctuations, and scale-dependent anisotropies,''
  Phys.\ Rev.\ D {\bf 84} (2011) 063528
  [arXiv:1104.0002 [hep-th]].


\bibitem{Gerstenlauer:2011ti}
  M.~Gerstenlauer, A.~Hebecker and G.~Tasinato,
  ``Inflationary Correlation Functions without Infrared Divergences,''
  JCAP {\bf 1106} (2011) 021
  [arXiv:1102.0560 [astro-ph.CO]].


\bibitem{DGLAP}
 D.~Seery,
  ``A parton picture of de Sitter space during
  slow-roll inflation,''
  JCAP {\bf 0905}, 021 (2009)
  [arXiv:0903.2788 [astro-ph.CO]].

\bibitem{AB}
 N.~Afshordi and R.~H.~Brandenberger,
  ``Super-Hubble nonlinear perturbations during inflation,''
  Phys.\ Rev.\  D {\bf 63}, 123505 (2001)
  [arXiv:gr-qc/0011075].

\bibitem{LU}
  B.~Losic and W.~G.~Unruh,
  ``Cosmological Perturbation Theory in Slow-Roll Spacetimes,''
  Phys.\ Rev.\ Lett.\  {\bf 101}, 111101 (2008)
  [arXiv:0804.4296 [gr-qc]].


\bibitem{JMPW}
 T.~M.~Janssen, S.~P.~Miao, T.~Prokopec and R.~P.~Woodard,
  ``Infrared Propagator Corrections for Constant Deceleration,''
  Class.\ Quant.\ Grav.\  {\bf 25} (2008) 245013
  [arXiv:0808.2449 [gr-qc]].

\bibitem{deSRG}
 C.~P.~Burgess, L.~Leblond, R.~Holman and S.~Shandera,
  ``Super-Hubble de Sitter Fluctuations and the Dynamical RG,''
  JCAP {\bf 1003} (2010) 033
  [arXiv:0912.1608 [hep-th]];
%
  ``Breakdown of Semiclassical Methods in de Sitter Space,''
  JCAP {\bf 1010} (2010) 017
  [arXiv:1005.3551 [hep-th]].

\bibitem{CMBthn}
 L.~H.~Ford,
  ``Quantum Instability Of De Sitter Space-Time,''
  Phys.\ Rev.\  D {\bf 31} (1985) 710;

 V.~Muller, H.~J.~Schmidt and A.~A.~Starobinsky,
  ``The Stability Of The De Sitter Space-Time
  In Fourth Order Gravity,''
  Phys.\ Lett.\  B {\bf 202} (1988) 198;

 A.~A.~Starobinsky,
  ``Stochastic de Sitter (inflationary)
  stage in the early universe,''
 {\it  In *De Vega, H.j. ( Ed.), Sanchez, N. ( Ed.): Field Theory,
 Quantum Gravity and Strings*, 107-126};

 I.~Antoniadis and E.~Mottola,
  ``Graviton Fluctuations In De Sitter Space,''
  J.\ Math.\ Phys.\  {\bf 32} (1991) 1037;

  M.~Sasaki, H.~Suzuki, K.~Yamamoto and J.~Yokoyama,
  ``Superexpansionary divergence: Breakdown of perturbative quantum field
  theory in space-time with accelerated expansion,''
  Class.\ Quant.\ Grav.\  {\bf 10} (1993) L55;

  A.~D.~Dolgov, M.~B.~Einhorn and V.~I.~Zakharov,
  ``On Infrared Effects In De Sitter Background,''
  Phys.\ Rev.\  D {\bf 52} (1995) 717
  [arXiv:gr-qc/9403056];


\bibitem{starobinsky}
  A.~A.~Starobinsky,
  ``Stochastic De Sitter (inflationary) Stage In The Early Universe,''
  Lect.\ Notes Phys.\  {\bf 246} (1986) 107;

 A.~A.~Starobinsky and J.~Yokoyama,
  ``Equilibrium state of a selfinteracting scalar field in the De Sitter background,''
  Phys.\ Rev.\ D {\bf 50} (1994) 6357
  [astro-ph/9407016].

\bibitem{NonPLdecoherence}
  M.~A.~Sakagami,
  ``Evolution From Pure States Into Mixed States In De Sitter Space,''
  Prog.\ Theor.\ Phys.\  {\bf 79}, 442 (1988);

  L.~P.~Grishchuk and Y.~V.~Sidorov,
  ``On The Quantum State Of Relic Gravitons,''
  Class.\ Quant.\ Grav.\  {\bf 6} (1989) L161;

  R.~H.~Brandenberger, R.~Laflamme and M.~Mijic,
  ``Classical Perturbations From Decoherence Of Quantum Fluctuations In The Inflationary Universe,''
  Mod.\ Phys.\ Lett.\ A {\bf 5}, 2311 (1990);

  E.~Calzetta and B.~L.~Hu,
  ``Quantum fluctuations, decoherence of the mean field, and structure formation in the early universe,''
  Phys.\ Rev.\ D {\bf 52}, 6770 (1995)
  [gr-qc/9505046];

  J.~Lesgourgues, D.~Polarski and A.~A.~Starobinsky,
  ``Quantum-to-classical transition of cosmological perturbations for  non-vacuum initial states,''
  Nucl.\ Phys.\ B {\bf 497}, 479 (1997)
  [gr-qc/9611019];

  C.~Kiefer, D.~Polarski and A.~A.~Starobinsky,
  ``Quantum-to-classical transition for fluctuations in the early universe,''
    Int.\ J.\ Mod.\ Phys.\ D {\bf 7}, 455 (1998)
  [gr-qc/9802003];

  C.~Kiefer and D.~Polarski,
  ``Emergence of classicality for primordial fluctuations: Concepts and analogies,''
  Annalen Phys.\  {\bf 7}, 137 (1998)
  [gr-qc/9805014];

  F.~C.~Lombardo and D.~Lopez Nacir,
  ``Decoherence during inflation: The generation of classical
  inhomogeneities,''
  Phys.\ Rev.\ D {\bf 72}, 063506 (2005)
  [gr-qc/0506051];

  J.~W.~Sharman and G.~D.~Moore,
  ``Decoherence due to the Horizon after Inflation,''
  JCAP {\bf 0711} (2007) 020
  [arXiv:0708.3353 [gr-qc]].


\bibitem{Guth:1985ya}
  A.~H.~Guth and S.~Y.~Pi,
  ``The Quantum Mechanics Of The Scalar Field In The New Inflationary
  Phys.\ Rev.\ D {\bf 32}, 1899 (1985).

\bibitem{Polarski:1995jg}
  D.~Polarski and A.~A.~Starobinsky,
  ``Semiclassicality and decoherence of cosmological perturbations,''
  Class.\ Quant.\ Grav.\  {\bf 13}, 377 (1996)
  [gr-qc/9504030].

\bibitem{squeezed}
 L.~P.~Grishchuk and Y.~V.~Sidorov,
 ``Squeezed Quantum States Of Relic Gravitons And Primordial Density
 Fluctuations,''
 Phys.\ Rev.\ D {\bf 42}, 3413 (1990);

  A.~Albrecht, P.~Ferreira, M.~Joyce and T.~Prokopec,
  ``Inflation and squeezed quantum states,''
  Phys.\ Rev.\ D {\bf 50}, 4807 (1994)
  [astro-ph/9303001].

\bibitem{OtherNoiseDeriv}
  S.~Habib,
  ``Stochastic inflation: The Quantum phase space approach,''
  Phys.\ Rev.\ D {\bf 46} (1992) 2408
  [gr-qc/9208006];

  J.~Weenink and T.~Prokopec,
  ``On decoherence of cosmological perturbations and stochastic inflation,''
  arXiv:1108.3994 [gr-qc];

  L.~Perreault Levasseur,
  ``Lagrangian formulation of stochastic inflation: Langevin equations, one-loop corrections and a proposed recursive approach,''
  Phys.\ Rev.\ D {\bf 88} (2013) 8,  083537
  [arXiv:1304.6408 [hep-th]].

\bibitem{Companion}
 ``Open EFTs: Effective Field Theories Without Effective Lagrangians,'' (in preparation).

\bibitem{Lindblad}
 C. Cohen-Tannoudji, J. Dupont-Roc and G. Grynberg, {\it Atom
 Photon Interactions}, Wiley, New York, 1992;

 V.F. Sears, {\it Neutron Optics}, Oxford University Press, 1989;

 H.~Haken,
 \textit{The Semiclassical and Quantum Theory of the Laser}, in
 \textit{Quantum Optics: Proceedings of the Tenth Session of the
 Scottish Universities Summer School in Physics, 1969} ed. by
 S.~M.~ Kay and A.~ Maitland, and references therein.

\bibitem{StochInfOld}
  C.~P.~Burgess, R.~Holman and D.~Hoover,
  ``Decoherence of inflationary primordial fluctuations,''
  Phys.\ Rev.\ D {\bf 77} (2008) 063534
  [astro-ph/0601646].

\bibitem{Kiefer:2006je}
  C.~Kiefer, I.~Lohmar, D.~Polarski and A.~A.~Starobinsky,
  ``Pointer states for primordial fluctuations in inflationary cosmology,''
  Class.\ Quant.\ Grav.\  {\bf 24} (2007) 1699
  [astro-ph/0610700].


\bibitem{BM}
 C.~P.~Burgess and D.~Michaud,
 ``Neutrino propagation in a fluctuating sun,''
 Annals Phys.\  {\bf 256}, 1 (1997) [hep-ph/9606295].

\bibitem{Feynman}
 R.~P.~Feynman and F.L. Vernon, jr.,
{\em Ann. Phys. (N.Y.)} 24, 118 (1963).

\bibitem{others}
 S.~Chaturvedy and F.~Shibata, Z. Phys.\ B35, 297;
 (1979);
%
  J.~R.~Anglin and W.~H.~Zurek,
  ``Decoherence of Quantum Fields: Pointer States and Predictability,''
  Phys.\ Rev.\ D {\bf 53}, 7327 (1996)
  [quant-ph/9510021].


\bibitem{Mukhanov}

  M.~Sasaki,
  ``Large Scale Quantum Fluctuations in the Inflationary Universe,''
  Prog.\ Theor.\ Phys.\  {\bf 76} (1986) 1036.


 V.~F.~Mukhanov,
 ``Quantum Theory Of Gauge Invariant Cosmological Perturbations,''
 Sov.\ Phys.\ JETP {\bf 67}, 1297 (1988)
 [Zh.\ Eksp.\ Teor.\ Fiz.\  {\bf 94N7}, 1 (1988)];

 V.~F.~Mukhanov, H.~A.~Feldman and R.~H.~Brandenberger,
 ``Theory Of Cosmological Perturbations. Part 1. Classical Perturbations. Part 2. Quantum Theory Of Perturbations. Part 3. Extensions,''
 Phys.\ Rept.\  {\bf 215}, 203 (1992).

 R.~H.~Brandenberger,
 ``Lectures on the theory of cosmological perturbations,''
 Lect.\ Notes Phys.\  {\bf 646}, 127 (2004) [hep-th/0306071].


\bibitem{dSWF}
C. Schomblond and P. Spindel, Ann.\ Inst.\ Henri Poincar\'e 25A
(1976) 67;
%
T.S.~Bunch and P.C.W.~Davies, Proc.\ Roy.\ Soc.\ (London) A360
(1977) 117.

\bibitem{dSProps}
  P.~Candelas and D.~J.~Raine,
  Phys.\ Rev.\ D {\bf 12}, 965 (1975);
  %
C.P. Burgess and C.A. Lutken, Phys.\ Lett.\ B153 (1985) 137;

\bibitem{dSflucts}
  L.~H.~Ford and A.~Vilenkin,
  ``Global Symmetry Breaking in Two-dimensional Flat Space-time and in De Sitter Space-time,''
  Phys.\ Rev.\ D {\bf 33} (1986) 2833.

\bibitem{Boyanovsky:1993xf}
  D.~Boyanovsky, H.~J.~de Vega and R.~Holman,
  ``Nonequilibrium evolution of scalar fields in FRW cosmologies I,''
  Phys.\ Rev.\ D {\bf 49} (1994) 2769
  [hep-ph/9310319].

\bibitem{Finelli}
  F.~Finelli, G.~Marozzi, A.A.~Starobinsky, G.P.~Vacca and G.~Venturi, ``Stochastic growth of quantum fluctuations during slow-roll inflation,''
   Phys.\ Rev.\ D {\bf82} (2010) 064020, 
  [arXiv:1003.1327].

\bibitem{nufluct}
 P.~Bamert, C.~P.~Burgess and D.~Michaud,
  ``Neutrino propagation through helioseismic waves,''
  Nucl.\ Phys.\ B {\bf 513} (1998) 319
  [hep-ph/9707542].

\bibitem{TsamisWoodard}
N.~C.~Tsamis and R.~P.~Woodard,
  ``Stochastic quantum gravitational inflation,''
  Nucl.\ Phys.\ B {\bf 724} (2005) 295
  [gr-qc/0505115].

\end{thebibliography}
\end{document}